\documentclass[10pt,british,sigplan,screen,balance=true,authorversion=true]{acmart}
\pdfoutput=1
\usepackage[T1]{fontenc}
\usepackage{textcomp}
\usepackage[latin9]{inputenc}
\usepackage{color}
\usepackage{cprotect}
\usepackage{refstyle}
\usepackage{mathtools}
\usepackage{algorithm2e}
\usepackage{amssymb}

\makeatletter

\global\long\def\emptytex{}%
\global\long\def\b#1{\textcolor{black}{#1}}%
\global\long\def\a#1{\texttt{\textlangle}\b{#1}\texttt{\textrangle}}%
\global\long\def\p#1{\texttt{(}\b{#1}\texttt{)}}%
\global\long\def\k#1{\textcolor{blue}{\texttt{#1}}}%
\global\long\def\t#1{\textcolor{teal}{#1}}%
\global\long\def\f#1{\textcolor{purple}{#1}}%
\global\long\def\l#1{\textcolor{brown}{#1}}%
\global\long\def\v#1{\textcolor{orange}{#1}}%
\global\long\def\M#1{\textcolor{green}{#1}}%
\global\long\def\c#1{\texttt{#1}}%
\global\long\def\ck#1{\c{\k{#1}}}%
\global\long\def\ct#1{\c{\t{#1}}}%
\global\long\def\cf#1{\c{\f{#1}}}%
\global\long\def\cv#1{\c{\v{#1}}}%
\global\long\def\cM#1{\c{\M{#1}}}%
\global\long\def\cl#1{\c{\l{#1}}}%
\global\long\def\cb#1{\c{\b{#1}}}%
\global\long\def\mt#1{\ensuremath{\t{#1}}}%
\global\long\def\mf#1{\ensuremath{\f{#1}}}%
\global\long\def\mv#1{\ensuremath{\v{#1}}}%
\global\long\def\*{\cf *}%
\global\long\def\ob#1{\ensuremath{\overline{#1}}}%
\global\long\def\emp{\l{\upepsilon}}%
\global\long\def\everydisplaystyle#1{{\displaystyle #1}}%
\global\long\def\rn#1{\textsc{(#1)}}%
\global\long\def\nir#1#2#3{\everydisplaystyle{\rn{#1}\ \frac{\everydisplaystyle{#2}}{\everydisplaystyle{#3}}}}%
\global\long\def\op#1{\textrm{\ensuremath{\mathrm{#1}}}}%
\global\long\def\xs#1{\mv{x#1}}%
\global\long\def\x{\xs{\emptytex}}%
\global\long\def\ff{\mv f}%
\global\long\def\D{\mt D}%
\global\long\def\X{\mt{\ensuremath{X}}}%
\global\long\def\ees#1{\mf{e#1}}%
\global\long\def\ee{\ees{\emptytex}}%
\global\long\def\SIG{\mf S}%
\global\long\def\GG{\mv{\varGamma}}%
\global\long\def\DL{\mt{\varDelta}}%
\global\long\def\SI{\ensuremath{\mt{\varSigma}}}%
\global\long\def\co{\cb :\,}%
\global\long\def\eq{\cb{\,=\,}}%
\global\long\def\br#1{\texttt{\{}#1\texttt{\}}}%
\global\long\def\mm{\mf{\mu}}%
\global\long\def\ms#1{\mf{m#1}}%
\global\long\def\m{\ms{\emptytex}}%
\global\long\def\e#1{\ck{effect[}#1\ck ]}%
\global\long\def\re#1{\ck{restrict[}#1\ck ]}%
\global\long\def\eps#1{\mf{\ensuremath{\varepsilon}#1}}%
\global\long\def\ep{\eps{\emptytex}}%
\global\long\def\mma#1{\mf{\mu\a{#1}}}%
\global\long\def\CCp#1{\mt{C\p{#1}}}%
\global\long\def\CCs#1{\mt{C#1}}%
\global\long\def\CC{\CCs{\emptytex}}%
\global\long\def\sc{\cb ;}%
\global\long\def\ddd{}%
\global\long\def\mmap#1#2{\mf{\mu\a{#1}\p{#2}}}%
\global\long\def\Ts#1{\mt{T#1}}%
\global\long\def\T{\Ts{\emptytex}}%
\global\long\def\mp#1{\mf{m\p{\b{#1}}}}%
\global\long\def\vsep{\text{\textbackslash\textbackslash}}%
\global\long\def\cor{\ensuremath{\mathrel{:}}}%
\global\long\def\mms#1{\mf{\mu#1}}%
\global\long\def\Xs#1{\mt{X#1}}%
\global\long\def\SIGs#1{\mf{\SIG#1}}%
\global\long\def\dddc{\c{\ddd,}}%
\global\long\def\cddd{\c{,\ddd}}%
\global\long\def\cldotsc{,\ldots,}%
\global\long\def\QED{\hfill\blacksquare}%
\global\long\def\hsep{\hspace{0.75em}}%

\global\long\def\name{\text{Call}\Eulerconst}%
\global\long\def\rname{\text{Indirect-}\name}%
\global\long\def\rc#1#2{\begin{array}[b]{l}
\boxed{#1}\\
\begin{array}[b]{c}
\vphantom{#2}\\
\vphantom{{\emptytex\atop \emptytex}}
\end{array}
\end{array}}%
\global\long\def\frc#1{\begin{array}[b]{c}
\vphantom{{\emptytex\atop \emptytex}}\\
\vphantom{#1}\\
\vphantom{{\emptytex\atop \emptytex}}
\end{array}}%

\setcopyright{acmlicensed}
\acmPrice{15.00}
\acmDOI{10.1145/3359591.3359731}
\acmYear{2019}
\copyrightyear{2019}
\acmISBN{978-1-4503-6995-4/19/10}
\acmConference[Onward! '19]{Proceedings of the 2019 ACM SIGPLAN International Symposium on New Ideas, New Paradigms, and Reflections on Programming and Software}{October 23--24, 2019}{Athens, Greece}
\acmBooktitle{Proceedings of the 2019 ACM SIGPLAN International Symposium on New Ideas, New Paradigms, and Reflections on Programming and Software (Onward! '19), October 23--24, 2019, Athens, Greece} 

\AtBeginDocument{\providecommand\secref[1]{\ref{sec:#1}}}
\AtBeginDocument{\providecommand\subsecref[1]{\ref{subsec:#1}}}
\AtBeginDocument{\providecommand\Secref[1]{\ref{Sec:#1}}}
\AtBeginDocument{\providecommand\figref[1]{\ref{fig:#1}}}
\AtBeginDocument{\providecommand\Figref[1]{\ref{Fig:#1}}}
\RS@ifundefined{subsecref}
  {\newref{subsec}{name = \RSsectxt}}
  {}
\RS@ifundefined{thmref}
  {\def\RSthmtxt{theorem~}\newref{thm}{name = \RSthmtxt}}
  {}
\RS@ifundefined{lemref}
  {\def\RSlemtxt{lemma~}\newref{lem}{name = \RSlemtxt}}
  {}

\makeatletter
\newcommand\ShowInterColumnFrame{
	\usepackage{etoolbox}
	\geometry{showframe}
	\patchcmd{\@outputdblcol}
	  {{\normalcolor\vrule\@width\columnseprule}}
		{\hspace{-6pt}\vrule\@width0.4pt\hspace{6pt}\hfil%
	    {\normalcolor\vrule\@width\columnseprule}%
    	\hfil\rule{6pt}{0pt}\vrule\@width0.4pt\hspace{-6pt}}{}{}}
\makeatother

\usepackage[british]{babel}
\definecolor{blue}{HTML}{0000F0} %
\definecolor{purple}{HTML}{700090}
\definecolor{orange}{HTML}{F07000}
\definecolor{teal}{HTML}{0090B0}
\definecolor{brown}{HTML}{A00000} 
\definecolor{green}{HTML}{008000}
\definecolor{pink}{HTML}{F000F0} 

\input{listings-style}

\newcommand{\keystyle}{\color{blue}}
\newcommand{\methstyle}{\color{purple}}

\newcommand{\litstyle}{\color{brown}}

\DeclareFontEncoding{LS1}{}{}
\DeclareFontSubstitution{LS1}{stix}{m}{n}
\DeclareSymbolFont{STIXE}{LS1}{stix}{m}{n}
\DeclareSymbolFont{STIXBE}{LS1}{stix}{b}{n}
\DeclareMathSymbol{\NEulerconst}{\mathord}{STIXE}{"BA}
\DeclareMathSymbol{\BEulerconst}{\mathord}{STIXBE}{"BA}
\makeatletter
\def\b@series{b}
\newcommand\Eulerconst{\ifx\f@series\b@series \BEulerconst \else \NEulerconst \fi}
\makeatother

\newcommand\stary{\methstyle{*}}

\usepackage{tikz}
\makeatletter
\global\let\tikz@ensure@dollar@catcode=\relax
\makeatother

\newcommand\textlrangle[4]{%
	\colorlet{orig}{.}%
	\makebox[0pt][l]{\hspace{#2em}\raisebox{#3em}{%
	\begin{tikzpicture}[x=1em, y=1em]%
		\draw[fill=orig, draw=none] #1--cycle;%
	\end{tikzpicture}}}\phantom{#4}}

\renewcommand\textlangle{\textlrangle{(0.2532, 0.8396)--(0.3013, 0.8102)--(0.0700, 0.4054)--(0.3005, 0.0321)--(0.2541, 0.0000)--(0.0000, 0.3944)--(0.0000, 0.4174)}{0.1391}{-0.1731}{(}}
\renewcommand\textrangle{\textlrangle{(0.0481, 0.8396)--(0.0000, 0.8102)--(0.2313, 0.4054)--(0.0008, 0.0321)--(0.0472, 0.0000)--(0.3013, 0.3944)--(0.3013, 0.4174)}{0.0487}{-0.1731}{)}}

\renewcommand{\a}[1]{\texttt{\textlangle}\b{#1}\texttt{\textrangle}}

\renewcommand\ddd{\hbox to \fontcharwd\font`x{\hss.\hss\hss.\hss\hss.\hss}}
\renewcommand\ldots{\ensuremath{\text{\textcolor{black}{\rmfamily \ddd}}}}

\renewcommand{\op}[1]{\qopname\relax o{#1}}
\renewcommand{\everydisplaystyle}[1]{{\everymath={\displaystyle}\ensuremath{\displaystyle#1}}}

\renewcommand{\ob}[1]{\colorlet{orig}{.}\b{\ensuremath{\overline{\textcolor{orig}{#1}}}}}

\newcommand{\defActiveMathChar}[2]{%
        \begingroup\lccode`~=`#1\relax%
        \lowercase{\endgroup\def~}{#2}%
        \AtBeginDocument{\mathcode`#1="8000}%
}
\defActiveMathChar{.}{\mspace{-4mu}\texttt{.}\mspace{-2mu}}

\newcommand\fixmathh[2]{%
	\expandafter\let\csname old#2\endcsname=#1%
	\renewcommand{#1}{\ensuremath{\csname old#2\endcsname}}%
}
\newcommand{\fixmath}[1]{%
	\expandafter\fixmathh\csname#1\endcsname{#1}
}

\newlength{\rch}
\newcommand{\rcBox}[2]{\\[-0.5\rch]\boxed{\ \vphantom{\overline{\strut}}\smash{#2}\ }\\}
\newcommand{\rcDepth}[1]{\setbox0=\hbox{\vphantom{#1}}\dp0=\dimexpr\dp0+\rch\box0}
\newcommand{\rcGeneral}[2]{\rlap{
	\arraycolsep=0pt%
	\def\arraystretch{1}%
	\setlength\fboxsep{0pt}%
	\setlength{\rch}{\baselineskip}%
	\hspace{0.5\rch}%
	\begin{array}[b]{l}\ensuremath{#1}#2\end{array}}}

\renewcommand{\rc}[2]{\rcGeneral{\rcBox{}{#1}}{\rcDepth{#2}}}

\renewcommand{\frc}[1]{\rcGeneral{\\}{\rcDepth{#1}}}

\fixmath{vdash}
\fixmath{leq}
\fixmath{vartriangleleft}
\fixmath{preceq}
\fixmath{overline}
\fixmath{cong}
\fixmath{notin}

\newlength{\ruleW}

\renewcommand{\vsep}{\setbox0=\hbox{}\dp0=0.5em\relax\box0}

\newcommand{\rlrule}[4]{
	\settowidth{\ruleW}{#1}

	\addtolength{\ruleW}{((\textwidth-(\widthof{#3}+\ruleW))/(#2 + #4))*#2}}

\newcommand{\rrule}[4]{\rlrule{#4}{#2}{#3}{#1}\hbox{}\hfill\rule{\ruleW}{0.4pt}\vspace{-\baselineskip}\vspace{-1.4pt}}
\newcommand{\lrule}[4]{\rlrule{#3}{#1}{#4}{#2}\vspace{-1.4pt}\rule{\ruleW}{0.4pt}\hfill\hbox{}}

\let\omaketitle=\maketitle
\def\maketitle{\let\maketitle=\omaketitle}

\renewcommand\secref[2][]{Section#1 \ref{sec:#2}}
\renewcommand\subsecref[1]{Section \ref{subsec:#1}}
\renewcommand\Secref[2][]{Section#1 \ref{sec:#2}}
\renewcommand\figref[1]{Figure \ref{fig:#1}}
\renewcommand\Figref[1]{Figure \ref{fig:#1}}

\newlength\bh
\renewcommand\QED{%
	\settoheight\bh{\ensuremath{\Box}}%
	\hfill\fboxsep=0pt\colorbox{black}{%
		\hbox to \bh{\hss\ensuremath{\phantom{\Box}\hss}}}}

\usepackage{caption}

\newcommand{\hidehyphens}{ 
        \renewcommand{\lstlistingname}{Listing}
        \renewcommand{\bibliography}[1]{}
        \renewcommand{\texttt}[1]{}
        \excludecomment{lstlisting}
        \let\endlstlisting\relax
}

\makeatother

\usepackage{babel}
\usepackage{listings}
\renewcommand{\lstlistingname}{Listing}

\begin{document}
\title{$\name$: An Effect System for Method Calls}

\author{Isaac Oscar Gariano}
\affiliation{\institution{Victoria University of Wellington}\city{Wellington}\country{New Zealand}}
\email{Isaac@ecs.vuw.ac.nz}
\author{James Noble}
\affiliation{\institution{Victoria University of Wellington}\city{Wellington}\country{New Zealand}}
\email{kjx@ecs.vuw.ac.nz}
\author{Marco Servetto}
\affiliation{\institution{Victoria University of Wellington}\city{Wellington}\country{New Zealand}}
\email{Marco.Servetto@ecs.vuw.ac.nz}
\ccsdesc[500]{Software and its engineering~Object oriented languages}

\ccsdesc[500]{Software and its engineering~Language features}

\ccsdesc[300]{Software and its engineering~Imperative languages}

\ccsdesc[300]{Theory of computation~Type structures}

\keywords{object oriented languages, type and effect systems, side effects,
static type systems}\maketitle

\begin{abstract}
Effect systems are used to statically reason about the \emph{effects}
an expression may have when evaluated. In the literature, such effects
include various behaviours as diverse as memory accesses and exception
throwing. Here we present $\name$, an object-oriented language that
takes a flexible approach where effects are just \emph{method calls}:
this works well because ordinary methods often model things like I/O
operations, access to global state, or primitive language operations
such as thread creation. $\name$ supports both flexible and fine-grained
control over such behaviour, in a way designed to minimise the complexity
of annotations.

$\name$'s effect system can be used to prevent OO code from performing
privileged operations, such as querying a database, modifying GUI
widgets, exiting the program, or performing network communication.
It can also be used to ensure determinism, by preventing methods from
(indirectly) calling non-deterministic primitives like random number
generation or file reading.
\end{abstract}
\maketitle
\pagebreak{}

\section{Introduction}\label{sec:Introduction}

\emph{Type and effect systems }(or just \emph{effect systems})\emph{
}were originally introduced to reason about the purity of code in
functional programming languages \cite{Integrating_Functional_and_Imperative_Programming};
since then they have been applied to reason about many other properties
of code such as memory accesses \cite{Type_and_Effect_Systems,Polymorphic_Type_Region_and_Effect_Inference}
and exception throwing \cite{The_Java_Language_Specification_Third_Edition}.
Such pre-existing effect systems typically only have a small predefined
set of effects, such as $\ck{init(}\v{\rho}\ck )$/$\ck{read(}\v{\rho}\ck )$/$\ck{write(}\v{\rho}\ck )$
(where $\v{\rho}$ is a memory region variable), or $\c{\k{throws}}$
$\ob{\T}$ (where $\ob{\T}$ is a list of exception types) \cite{The_Java_Language_Specification_Third_Edition}.
Other limitations that have been identified are that they require
complex and verbose source code annotations in order for library code
to be useful \cite{Lightweight_Polymorphic_Effects}.

Here we focus on the general problem of restricting the use of library
defined \emph{effectful} operations: such as I/O or (indirect) access
to private global state. We present an OO language, $\name$\footnote{$\name$ is pronounced like `callee'; the $\Eulerconst$ stands
for `effect'.}, which uses an effect system that works neatly with the OO concepts
of sub-typing and generics, without requiring separate features like
effect polymorphism \cite{Polymorphic_Effect_Systems}.

Like other effect systems, $\name$ works by typing each expression
with a list of \emph{effects}: here effects are \emph{method names},
indicating \emph{behaviour} that may occur at runtime; the effects
of an \emph{expression} are simply the names of all the methods it
\emph{directly }calls in any sub-expression. Thus performing an \emph{effect}
simply corresponds to calling a method: what this means is up to the
individual method in question.

Methods must be declared with an $\e{\ob{\ep}}$\footnote{We use the notation $\ob{\ep}$ to represent a list $\eps{_{1}}\cldotsc\eps{_{n}}$,
for some $n\geq0$.} annotation, where the list of effects $\ob{\ep}$ represents an \emph{upper}-bound
on the method's behaviour: its body can only have the behaviour of
an expression with effects $\ob{\ep}$. This \emph{doesn't} mean that
the list of methods it calls is exactly $\ob{\ep}$, rather we allow
the body to have any \emph{sub-effect} of $\ob{\ep}$. In particular,
a method declared with (a sub-effect of) the empty list is \emph{uneffectful}:
it can be called by any method (including other uneffectful ones).
Other methods are however \emph{effectful}, and can only be called
by methods with a sufficiently strong $\ck{effect}$ annotation.

$\name$ is designed to minimise the complexity of such annotations,
whilst still allowing useful reasoning as to when effectful methods
can be called; the core mechanism by which we do this is the following
novel rule:

\vspace{1.5\smallskipamount}

\textbf{The $\rname$ Rule:} \emph{If a method $\T.\m$ is annotated
with $\e{\ob{\ep}}$, then $\T.\m$ is a sub-effect of $\,\ob{\ep}$.
In particular, $\T.\m$ can be called by any method annotated with
(super-effects of) $\,\ob{\ep}$.}

\vspace{1.5\smallskipamount}

This rule makes sense because $\T.\m$ can only have the behaviour
allowed by $\ob{\ep}$. Note that the inverse does not hold: it is
unlikely that one can perform the equivalent of arbitrary calls to
methods in $\ob{\ep}$ by simply calling $\T.\m$.

The following example library for interacting with the console illustrates
the core parts of $\name$\footnote{$\name$ is an expression based language, which uses semicolons as
the conventional sequencing operator.}:

\noindent 
\begin{lstlisting}
class Console {
	// Trusted I/O method that prints to the console,
	// defined in a possibly different language.
	foreign static Void #print(Char c) effect[#print];

	// Utility method to print a whole string at a time!
	static Void #print-str(String s) effect[#print] =
		for (Char c in s) #print(c);

	// Prints a line
	static Void #print-line(String s) effect[#print] =
		#print-str(s + "\n");

	// Reads a single character from the console
	foreign static Char #read() effect[#read];
	... }
\end{lstlisting}
Here $\cf{print}$ is declared with $\e{\cf{print}}$; this ensures
that the $\rname$ rule only applies to effects already containing
$\cf{print}$: only a method that lists $\cf{print}$ in its $\ck{effect}$
annotation can directly call it.\footnote{As $\ck{foreign}$ methods do not have a body which we can type-check,
we do not restrict their effect annotation. An implementation however
may wish to restrict this, such as by fixing the effect annotations
for $\ck{foreign}$ methods declared by untrusted source code.} Now we have $\cf{print-str}$ which is declared with $\e{\cf{print}}$:
meaning the only effectful behaviour it can perform is that of the
$\cf{print}$ method (the only effectful method called).\footnote{We assume that pure operations (such as iteration and addition) on
standard types (like $\ct{String}$ and $\ct{Int}$) are all declared
with $\e{\emptytex}$: i.e. they are uneffectful.} Now consider $\cf{print-line}$, instead of implementing it with
a loop like $\cf{print-str}$'s, the $\rname$ allows it to simply
call the $\cf{print-str}$ method. This demonstrates the main justification
behind the $\rname$: one can obtain the same behaviour as $\cf{print-str}$
by only calling $\cf{print}$ (together with uneffectful methods).
Note that none of these $\cf{print-}$ methods can (indirectly) call
$\cf{read}$, as $\cf{read}$ is not in their $\ck{effect}$ annotations,
neither is any method that (transitively) has the $\cf{read}$ effect.

Now consider the following example:
\begin{lstlisting}
static Void #hello() effect[Console.#print] = (
	restrict[] Untrusted.#untrusted();
	Console.#print-line("Hello World!"));
\end{lstlisting}
A $\re{\ob{\ep}}$ $\ee$ expression ensures that the effects of $\ee$
are sub-effects of $\ob{\ep}$. In the case above, since $\ob{\ep}=\emp$,
the type checker will check that the $\cf{untrusted}$ effect is a
sub-effect of the empty-effect, i.e. it is declared to not (transitively)
call any effectful methods. Assuming that all other I/O methods provided
by the standard library are effectful, we can be sure that the only
I/O $\cf{hello}$ will perform is to print `$\cl{Hello}$ $\cl{World!}$'.
Note that the $\rname$ rule allows $\cf{hello}$ to call $\cf{print-line}$,
just as it allowed $\cf{print-line}$ to call $\cf{print}$. 

The $\rname$ rule is particularly flexible as it allows for abstraction:
the implementation details of the $\cf{hello}$ method are not exposed
by its effect annotation (we know that it \emph{may} eventually call
$\cf{print}$, but not that it does so by calling $\cf{print-line}$).
This is particularly important if $\cf{print-line}$ were to be a
private method, then mentioning it in the effect annotation of $\cf{hello}$
would expose its existence to the public. Providing a more\emph{ }specific\emph{
}effect annotation than the $\rname$ rule requires can however allow
for stronger reasoning: if $\cf{hello}$ were to be annotated with
$\e{\cf{print-line}}$, we can be sure that everything it outputs
will be terminated by a newline. In order to perform such reasoning,
\emph{callers} of $\cf{hello}$ only need to look at the body of $\cf{print-line}$,
they need not look at the body of $\cf{hello}$ itself. This however
shows a disadvantage of our system: to modify $\cf{hello}$ to enable
such reasoning we have to look at its body and expose more implementation
details in its effect annotation.\footnote{Effect annotations are similar to return type specifications: one
can expose more implementation details by declaring more specific
return types. For example, consider $\ct{List}$ $\cf{foo()}$ $\c =$
$\ck{new}$ $\ct{Linked-List()}\c ;$ $\cf{foo}$ could expose more
implementation details by specifying $\ct{Linked-List}$ as its return
type, or it could expose less useful information by specifying $\ct{Object}$.}

This same line of reasoning we did on $\c{\t{Console}.\f{read}}$
works with any other effectful operations one can define as methods,
such as $\ct{File-System}\c .\cf{append}$, $\ct{Thread}\c .\cf{spawn}$,
$\ct{Socket}\c .\cf{write}$, $\ct{Program}\c .\cf{exit}$, or $\ct{Random}\c .\cf{generate}$
methods.

\subsection{Contents}

The rest of the paper is structured as follows:
\begin{itemize}
\item In \secref{OO} we explain how $\name$'s effect system works in the
context of OO, in particular dynamic dispatch and generics.
\item In \secref{Reasoning} we demonstrate how $\name$'s effects can be
used to reason about library defined behaviour, including operations
on global state, creation of new instances of a class, and restricted
forms of I/O.
\item In \secref{Formalism} we present a minimal grammar for $\name$ and
the typing rules for the novel parts of $\name$'s type system, in
particular the sub-effect relation. The remaining non-novel parts
of our type system are presented in Appendix \ref{sec:Auxiliary}.
We do not present a formalism of the runtime semantics, since they
are merely a sub-set of Featherweight Generic Java's \cite{Featherweight_Java:_A_Minimal_Core_Calculus_for_Java_and_GJ}.
\item In \secref{Extensions} we outline how we might add three additional
features to $\name$. In particular, we show how effects can be combined
with generic type parameter variance, dynamic code loading/invocation,
and method redirection.
\item In \secref{Example} we present two larger use cases for $\name$'s
effect system: preventing security vulnerabilities caused by database
accesses, and preventing untrusted advertisement code from modifying
parts of a GUI.
\item In \secref{Soundness} we informally make two statements of soundness,
and provide sketches for their proofs.
\item In \secref{Related} we discuss related work, and show how $\name$
compares.
\item Finally, in \secref{Conclusion} we summarise our results and conclude.
\end{itemize}

\section{Object-Orientation with Effects}\label{sec:OO}

\subsection{Dynamic Dispatch}\label{subsec:OO-Interface}

Like many other effect systems, a method's $\ck{effect}$ annotation
represents an \emph{upper }bound. In particular, they can list more
methods than are actually called. This works well together with sub-typing
and dynamic dispatch:

\begin{lstlisting}[escapechar={`}]
interface UI-Element {
	Void #paint() effect[UI.#set-pixel];
	List<UI-Element> #children() effect[];
	... }

static Void #paint-all(UI-Element e) effect[UI.#set-pixel] = (
	e.#paint(); // well typed, by the `$\textcolor{green}{\rname}$` rule
	for (UI-Element c in e.#children()) #paint-all(c)`\textcolor{black}{)}`;
\end{lstlisting}
As in the above $\cf{paint-all}$ method, the $\rname$ rule always
allow recursive calls, since they cannot be used to perform additional
effectful behaviour. From the above code, we know that the only effectful
behaviour $\cf{paint-all}$ can do is that of $\c{\t{UI}.\f{set-pixel}}$;
however what $\c{\t{UI}.\f{set-pixel}}$ calls will actually occur
is up to the implementation of $\cf{paint}$ for the specific $\ct{UI-Element}$s
given. For example, consider the following:

\begin{lstlisting}
class Empty-Element: UI-Element {
	Void #paint() effect[] = skip; // Don't do anything
	List<UI-Element> #children()  effect[] = {}; // Empty list
	... }
\end{lstlisting}
As with many other OO languages, we allow method types to be refined:
the effects in a method's $\ck{effect}$ annotation must be sub-effects
of those in the methods it is overriding. Thus the above code is valid
since $\c{\t{Empty-Element}.\f{paint}}$ is declared to have no effects,
which are vacuously sub-effects of $\c{\t{UI}.\f{set-pixel}}$.

\subsection{F-Bounded Polymorphism}

Many object-oriented languages support F-bounded polymorphism for
generic code, such as in the following typical use case:

\begin{lstlisting}
interface Hashable<T> {
	Bool #equals(T other);
	Int #hash(); }
class Hash-Map<Key: Hashable<Key>, Value> {
	// Needs to call $Key.#hash and $Key.#equals
	Value #get(Key k) = ... ;
	... }
\end{lstlisting}
But what $\ck{effect}$ annotation should the above methods have?
The $\ct{Hash-Map}$ and $\ct{Hashable}$ classes are likely to be
used by lots of very different code with different desires. For example,
some users of $\ct{HashMap}$ might want to ensure that accessing
elements won't perform any I/O, which some implementations of $\ct{Hashable}.\cf{hash}$
will perform. However, some users may not mind if I/O is performed,
and one may wish to implement $\ct{Hashable}.\cf{hash}$ using a hardware
random number generator. It may seem that one cannot provide a single
$\ct{Hash-Map}$ class that is simultaneously usable in these cases.
However, note how the $\ct{Hash-Map}$ class is already generic on
the $\ct{Key}$ type, different $\ct{Key}$ types can cause $\cf{get}$
to perform different behaviour at runtime. This means we need not
restrict the effects of the methods in $\ct{Hashable}$: rather we
can look at the effects for a given $\ct{Key}$. As such, we will
annotate the methods of $\ct{Hashable}$ with the special wildcard
effect $\*$, which is a super effect of all other effects (i.e. $\name$
does not restrict the behaviour/method calls of a method declared
with the $\*$ effect):

\begin{lstlisting}
interface Hashable<T> {
	Bool #equals(T other) effect[*];
	Int #hash() effect[*]; }
class Hash-Map<Key: Hashable<Key>, Value> {
	Value #get(Key k) effect[Key.#equals, Key.#hash] = ... ;
	... }
\end{lstlisting}
This is particularly neat as we do not require separate generic parameters
for effects, as is usual in effect system \cite{Koka:_Programming_with_Row-polymorphic_Effect_Types},
rather we can just use a normal type-parameter to abstract over effects.
In particular, note how the effect of $\cf{get}$ depends on a generic
type parameter: if we have more information about the $\ct{Key}$
type, we know more about $\cf{get}$:

\begin{lstlisting}
interface Random-Hashable<T>: Hashable<T> {
	Bool #equals(T other) effect[];
	Int #hash() effect[Random.#generate]; }

static Void #f<K: Random-Hashable<K>, V>(Hash-Map<K, V> m, K k)
	effect[Random.#generate] = m.#get(k);
\end{lstlisting}
The $\rname$ rule allows the last line to type check: we know that
any subtype of $\ct{Random-Hashable}$ must respect its interface,
in particular its $\cf{hash}$ and $\cf{equals}$ methods must be
declared with sub-effects of $\c{\t{Random}.\cf{generate}}$. Note
how even though we used the all-powerful $\*$ effect when declaring
$\ct{Hashable}$, we have not lost the ability to reason since the
effect can be arbitrarily refined by implementing classes/interfaces
(such as $\ct{Random-Hashable}$ above).

\section{Reasoning Power}\label{sec:Reasoning}

Here we present examples showing the kind of reasoning $\name$ allows.
In particular, we show how libraries can define new effectful methods
in order to reason about the behaviour of untrusted code, provided
that such code is type-checked.

\subsection{Indirect Effects}\label{subsec:Reasoning-Indirect}

Consider a library for operating on the file system and a $\cf{log}$
function that uses this library:

\begin{lstlisting}
class File-System {
	// Appends text to filename
	foreign static Void #append(String filename, String text)
		effect[#append]; 
	... }

static Void #log(String text) effect[File-System.#append] =
	File-System.#append("log.txt", text);

static Void #do-something() effect[#log] = ...;
\end{lstlisting}
Assuming that all methods that operate on the file system (such as
$\cf{append}$) are effectful, we can be sure that the only such operation
that $\cf{do-something}$ can perform is to append to $\cl{log.txt}$.
This only requires inspecting the code of $\cf{log}$ itself; the
body of $\cf{do-something}$ is irrelevant, since our effect system
will ensure that it can't call effectful operations (such as $\c{\t{File-System}.\f{append}}$)
unless their only possible effects are calls to $\cf{log}$. Note
how the effects of an expression are not the same as its \emph{indirect
effects}: $\cf{do-something}$ does \emph{not} have the $\c{\t{File-System}.\f{append}}$
effect, but it does indirectly, through the $\cf{log}$ effect.

In particular, the above example shows why it can be useful to declare
more specific effects: if the programmer doesn't mind if the log file
is appended, they can safely call $\cf{do-something}$, even if they
wouldn't trust $\cf{do-something}$ to append to \emph{arbitrary }files.
If a programmer is however unfamiliar with $\cf{log}$, if it is dynamically
dispatched, or if it has no source code to inspect, then the programmer
can look at its $\ck{effect}$ annotation to see that $\cf{do-something}$
might indirectly call $\c{\t{File-System}.\f{append}}$, but not $\ct{Console}\c .\cf{read}$.
A tool (such as an IDE) could help with such an investigation by providing
a graph of all (indirect) effect annotations; doing so only requires
that method signatures be available, not their bodies.

Alternatively, thanks to the $\rname$ rule, if there is a specific
set of effects that the programmer trusts, they can use our $\ck{restrict}$
expression (see \secref{Introduction}) without looking at any other
effect annotations.

\subsection{Object Creation}\label{subsec:Object-Creation}

Consider grating an untrusted component read access to files, but
only those in a runtime-determined list of files/folders. One way
to do this would be to give the component a list of paths to each
file/folder it can read; then by preventing the component from creating
new file paths, we can be sure that only those files are read:

\begin{lstlisting}
class Path { 
	// Converts 's' to a Path object
	static Path #parse(String s) effect[#parse] = ... ;
	// Appends the given file/directory name to the path
	Void #append-path(String suffix) effect[] = ...;
	
	// Reads the contents of the identified file
	String #read() effect[#read] = ...;

	// Assuming a private constructor
	... }

static Void #untrusted(List<Path> allowed-paths) 
	effect[Path.#read] = ...;
static Void #trusted() effect[Path.#parse, Path.#read] = 
	#untrusted({Path.#parse("~/"), Path.#parse("/tmp/")});
\end{lstlisting}
Since $\cf{untrusted}$ is not declared with the $\c{\t{Path}.\f{parse}}$
effect, we can be sure that it can't create new $\ct{Path}$ objects
out of thin air; rather it can only read from folders it obtains from
elsewhere. In the above call to it in $\cf{trusted}$, we can be sure
that it can only read from the user's home directory, or the system
wide $\cl{tmp}$ directory (this however does require the absence
of any other pre-existing $\ct{Path}$s accessible through global
variables).

\subsection{Global State}

As an $\ck{effect}$ annotation represents an upper bound, we can
provide effects which are stronger than necessary to type check the
methods body. This has two important advantages: method signatures
can be more resilient to implementation changes, and one can make
`privileged' methods harder to call. The second part is very useful:
it allows one to declare new \emph{effectful }operations, even if
their bodies are not otherwise effectful. Recall the UI example from
\subsecref{OO-Interface}, we mentioned a $\cf{set-pixel}$ function,
but what is it? It could be an externally defined method like our
$\cf{print}$ function from \secref{Introduction}. Alternatively,
we could implement it with a global variable, and delay updating the
actual display till later:
\begin{lstlisting}
class UI {
	// 'global' variable
	private static Array2D<Pixel> pixels = ...;
	Void #set-pixel(Pixel p, Int x, Int y) effect[#set-pixel] =
		// Assuming that set is not an effectful method
		pixels.#set(x, y, p);

	// The window will only update when this is called
	static Void #update-display() effect[#update-display] = 
		#set-window(pixels);

	// Defined externally, probably by the OS/Window Manager
	private foreign static Void #set-window(Array2D<Pixel> p)
		effect[];
	... }
\end{lstlisting}
Notice that we haven't marked the $\cf{set-window}$ function as effectful:
since it is $\ck{private}$, one merely needs to inspect the code
for the $\ct{UI}$ class to know that the only way for it to be called
is through the effectful $\cf{update-dislay}$ method. Since we do
not restrict the effect annotations of $\ck{foreign}$ methods, we
allow this, even though one may semantically consider $\cf{update-display}$
to be `effectful'. In fact, marking $\cf{set-window}$ as effectful
could break encapsulation, by requiring the signature of the public
$\cf{update-display}$ method to modify its $\ck{effect}$ annotation.
\begin{figure*}
{\footnotesize{}$\rc{\DL\vdash\ensuremath{\ob{\ep}\preceq\ob{\eps '}}}{\nir{\preceq Com}{\DL\vdash\ob{\eps{_{1}}}\preceq\ob{\eps{'_{1}}}\hsep\DL\vdash\ob{\eps{_{2}}}\preceq\ob{\eps{'_{2}}}}{\DL\vdash\ob{\eps{_{1}}},\ob{\eps{_{2}}}\preceq\DL\vdash\ob{\eps{'_{1}}},\ob{\eps{'_{2}}}}}$\hfill{}$\nir{\preceq Ref}{\bigl\{\ob{\ep}\bigr\}=\bigl\{\ob{\eps '}\bigr\}}{\DL\vdash\ob{\ep}\preceq\ob{\eps '}}$\hfill{}\hfill{}$\nir{\preceq Trans}{\DL\vdash\overline{\ep}\preceq\overline{\eps '}\hsep\DL\vdash\ob{\eps '}\preceq\overline{\eps{''}}}{\DL\vdash\ob{\ep}\preceq\ob{\eps{''}}}$\hfill{}\hfill{}$\nir{\preceq Comp}{\DL\vdash\ob{\eps{_{1}}}\preceq\ob{\eps{'_{1}}}\hsep\DL\vdash\ob{\eps{_{2}}}\preceq\ob{\eps{'_{2}}}}{\DL\vdash\ob{\eps{_{1}}},\ob{\eps{_{2}}}\preceq\DL\vdash\ob{\eps{'_{1}}},\ob{\eps{'_{2}}}}$\hfill{}\hbox{}}{\footnotesize\par}

{\footnotesize{}\hbox{}\hfill{}$\nir{\preceq\T}{\DL\vdash\T\leq\Ts '}{\DL\vdash\T.\m\preceq\Ts '.\m}$\hfill{}\hfill{}$\nir{\preceq\ensuremath{\CC.\m}}{\SI(\CC.\m)=\_\,\mmap{\_}{\_}\,\e{\ob{\ep}}\,\_\sc}{\DL\vdash\CC.\m\preceq\ob{\ep}}$\hfill{}\hfill{}$\nir{\preceq\*}{\emptytex}{\DL\vdash\overline{\ep}\preceq\*}$\hfill{}\hfill{}$\nir{\preceq\ensuremath{\emp}}{\emptytex}{\DL\vdash\emp\preceq\overline{\ep}}$\hfill{}\hbox{}}{\footnotesize\par}

\caption{Sub-effect Rules}\label{fig:Sub-effect-Rules}
\end{figure*}

Since $\cf{set-pixel}$ does not contain any calls to effectful methods,\footnote{For simplicity, $\name$ only treats method calls as effectful; if
we wished to restrict when $\k{static}$ variables can be read/written,
we could type such expression with effects such as $\k{static-read}$/$\k{static-write}$,
where for the purposes of the effect system, $\k{static-read}$/$\k{static-write}$
are the names of some predefined effectful methods.} we could have annotated it with $\e{\emptytex}$. By annotating it
with $\e{\cf{set-pixel}}$, we can now reason as to what methods can
call it. As in our $\cf{print}$ example from \secref{Introduction},
a method $\t T.\m$ can only \emph{directly }call $\cf{set-pixel}$
if $\t T.\m$ is annotated with $\e{\c{\dddc\f{set-pixel}\cddd}}$.
Similarly, $\t T.\m$ can \emph{indirectly} call $\cf{set-pixel}$
only if $\t T.\m$ is annotated with $\e{\c{\dddc\ensuremath{\Ts '.\ms '}\cddd}}$
and $\Ts '.\ms '$ can directly or indirectly call $\ck{\f{set-pixel}}$.
In particular, after checking that no other code in $\ct{UI}$ modifies
$\cv{pixels}$, we can be sure that a method call to $\t T.\m$ could
modify $\cv{pixels}$ only if the effect annotation of $\t T.\m$
allows it to indirectly call $\ck{\f{set-pixel}}$.

\section{Formalism}\label{sec:Formalism}

We will now present a formalism of $\name$, focusing on the novel
parts of our effect system. Our language is based on Featherweight
Generic Java (FGJ) \cite{Featherweight_Java:_A_Minimal_Core_Calculus_for_Java_and_GJ},
but in order to keep our formalism simple and our discussion focused
on our effect system, we have minimised the number of language features
we have formalised. In particular, we do not formalise generic classes
(only generic methods) or \emph{user defined} constructors. In addition,
we support interfaces and only a limited form of inheritance: classes
can inherit from multiple super interfaces, interfaces can not inherit
from other interfaces, and classes cannot be inherited. Though our
examples used such features, as well as additional ones such as accessibility
control, static methods, mutable state, and global variables, we do
not include them in our formalism as they do not make the novel parts
of our language more interesting.

In addition, we do not present reduction rules, as $\name$'s only
novel runtime expression ($\re{\ob{\ep}}\,\ee$) has a trivial reduction
(namely $\re{\ob{\ep}}\,\ee\to\ee$), rather we refer to FGJ for our
runtime semantics. Note that as is conventional with statically typed-languages,
we assume that the expression that is being reduced (i.e. the `main
expression') is initially well-typed; however we do not impose any
restriction on what such an expressions types or effects are.

\subsection{Grammar}\label{subsec:Grammar}

We will use the meta-variables $\x$, $\ff$, and $\mm$ as identifiers
for variables, fields, and methods (respectively). Similarly, $\CC$
and $\X$ identify classes/interfaces and generic type parameters,
respectively. The abstract syntax of $\name$ has the following grammar:\footnote{We use that notation $\ob{\X\co\CC}$ to mean $\Xs{_{1}}\co\CCs{_{1}}\cldotsc\Xs{_{n}}\co\CCs{_{n}}$,
for some $n\geq0$; similarly for the other overbars. Overbars can
also nest, for example $\ob{\mma{\ob{\T}}}$ is shorthand form $\mms{_{1}}\f{\a{\ob{\Ts{_{1}}}}}\cldotsc\mms{_{n}}\f{\a{\ob{\Ts{_{n}}}}}$,
which is itself shorthand for $\mms{_{1}}\f{\a{\Ts{_{1,1}}\cldotsc\Ts{_{1,k_{1}}}}}\cldotsc\mms{_{n}}\f{\a{\Ts{_{n,1}}\cldotsc\Ts{_{1,k_{n}}}}}$.
We use this interpretation since it is less ambiguous than the more
common $\ob{\X}\co\ob{\CC}$, which we would interpret as meaning
$\Xs{_{1}}\cldotsc\Xs{_{n}}\co\CCs{_{1}}\cldotsc\CCs{_{n}}$.}

{\footnotesize{}\vspace{1.5\smallskipamount}$\begin{aligned}\T\Coloneqq{\emptytex} & \CC\mid\X\\
\m\Coloneqq{\emptytex} & \mma{\ob{\T}}\\
\SIG\Coloneqq{\emptytex} & \T\,\mmap{\ob{\X\co\CC}}{\ob{\Ts '\,\x}}\,\e{\ob{\ep}}\\
\D\Coloneqq{\emptytex} & \k{class}\,\CCp{\ob{\CCs '\,\ff}}\co\ob{\CCs{''}}\,\br{\ob{\SIG\eq\ee\sc}}\mid\k{interface}\,\CC\,\br{\ob{\SIG\sc}}\\
\ee\Coloneqq{\emptytex} & \x\mid\ck{new}\,\CCp{\ob{\ee}}\mid\ee.\ff\mid\ee.\mp{\ob{\ees '}}\mid\re{\ob{\ep}}\,\ee\\
\ep\Coloneqq{\emptytex} & \*\mid\T.\m
\end{aligned}
$\medskip{}
}{\footnotesize\par}

\noindent A \emph{type}, $\T$, is either the name of a class or a
generic type parameter. A \emph{method selector}, $\m$, is a method
name $\mm$, provided with $\ob{\T}$ for its type parameters. A method
\emph{signature}, $\SIG$, specifies a return type $\T$, a method
name $\mm$, generic parameters $\ob{\X}$ (with upper bounds $\ob{\CC}$),
parameter names $\ob{\x}$ (with types $\ob{\Ts '}$), and a list
of effects $\ob{\ep}$. A (top-level) declaration, $\D$, either declares
a \emph{class} $\CC$ with fields $\ob{\ff}$ (of types $\ob{\CCs '}$),
implemented interfaces $\ob{\CCs{''}}$, and method implementations
with signatures $\ob{\SIG}$ and bodies $\ob{\ee}$; or declares an
\emph{interface} with name $\CC$ requiring the method signatures
$\ob{\SIG}$. An expression $\ee$ is either a variable name, a $\ck{new}$
expression, a field access, a method call, or an effect restriction.
Finally, an \emph{effect}, $\ep$, either names the wildcard effect
or a method.

Additionally, we will use $\GG:\x\rightarrow\T$ as a variable environment.
We will use $\DL:\X\rightarrow\CC$ as a mapping of generic parameters
to their upper bounds; to simplify our typing rules we will overload
notation and define $\DL(\CC)\coloneqq\CC$, for any $\CC$. We will
assume a global, and constant class table $\SI$, where $\SI(\CC)$
is the class or interface declaration for $\CC$. We assume the obvious
definition for $\SI(\CC.\mm)$ to extract the method implementation/specification
with name $\mm$ from the class/interface $\CC$ (i.e. $\SI(\CC.\mm)$
will be of form $\SIG\eq\ee$ or $\SIG$). We also use $\SI(\CC.\mma{\ob{\T}})$
to mean $\SI(\CC.\mm)[\ob{\X\coloneqq\T}]$, where $\ob{\X}$ are
the generic parameters declared by $\SI(\CC.\mm)$, except that this
substitution does not apply to the occurrences of $\X$ on the LHS
of a `$\c :$' (i.e. when $\X$ is in a binding position).

\subsection{The Sub-Effect Relation}

The core novelty of our type system is our sub-effect judgement, $\DL\vdash\ensuremath{\ob{\ep}\preceq\ob{\eps '}}$,
indicating that under the generic-parameter environment $\DL$, the
effects $\ob{\ep}$ are at most as effectful as $\ob{\eps '}$. \Secref[s]{Introduction}
\& \ref{sec:OO} informally describe when and why we allow this to
hold; however here in \figref{Sub-effect-Rules}\footnote{We use underscores to match against an arbitrary sequence of syntax
trees.} we formalise exactly when it holds. Our $\rn{\preceq Ref}$ rule
establishes $\preceq$ as reflexive, and that the order, and any duplication
of effects is irrelevant. Our $\rn{\preceq Trans}$ rule simply establishes
transitivity, and $\rn{\preceq Comp}$ establishes that composing
two sequences of effects preserves sub-effecting. The $\rn{\preceq\T}$
rule says that our sub-effect relation is covariant with respect to
sub-typing. Our $\rn{\preceq\ensuremath{\CC.\m}}$ rule formalises
our novel $\rname$ rule: the effects in the annotation of $\ensuremath{\T.\m}$
are super-effects of $\T.\m$. Finally our trivial $\rn{\preceq\*}$
and $\rn{\preceq\ensuremath{\emp}}$ rules simply establish $\*$
and $\emp$ as the greatest and least effects, respectively. Note
in particular that by applying $\rn{\preceq Ref}$, $\rn{\preceq Comp}$,
$\rn{\preceq\ensuremath{\emp}}$, and $\rn{\preceq Trans}$ rules
together, we have that if $\ob{\ep}$ is a subset of $\ob{\eps '}$,
then $\ob{\ep}$ is a sub-effect of $\ob{\eps '}$.

\subsection{Expression Typing}\label{subsec:Expression-Typing}

We use typing judgements of the form $\DL|\GG\vdash\ee:\T|\ob{\ep}$
to check that the expression $\ee$, using variables/type-parameters
in $\GG$/$\DL$ has type $\T$ and effects $\ob{\ep}$. Most of the
typing rules are pretty standard, and simply return the effects of
each sub-expression (if an expression has no sub-expressions, this
will be the empty list). However we have two non-standard rules: the
rule for method calls, and the rule for our novel restrict expression:{\scriptsize{}\vspace{0.5\baselineskip}}\\
{\scriptsize{}\hbox{}\hfill{}$\nir{\vdash\ensuremath{\ensuremath{\ee.\m}}}{\begin{array}{c}
\SI(\DL(\T)\,.\m)=\Ts{''}\,\mmap{\_}{\ob{\Ts '\,\x}}\,\_\sc\vsep\\
\DL\vdash\T.\m\hsep\DL|\GG\vdash\ee:\T|\ob{\ep}\hsep\ob{\DL|\GG\vdash\ees ':\Ts '|\ob{\eps '}}
\end{array}}{\DL|\GG\vdash\ee.\mp{\ob{\ees '}}:\Ts{''}|\ob{\ep},\ob{\ob{\eps '}},\T.\m}$\hfill{}\hbox{}}\\
{\scriptsize{}\vspace{\baselineskip}}\\
{\scriptsize{}\hbox{}\hfill{}$\nir{\vdash\textrm{\ensuremath{\ck{restrict}}}}{\DL|\GG\vdash\ee:\T|\ob{\eps '}\hsep\DL\vdash\ob{\eps '}\preceq\ob{\ep}}{\DL|\GG\vdash\re{\ob{\ep}}\,\ee:\T|\ob{\eps '}}$\hfill{}\hbox{}}\\

\noindent The $\rn{\vdash\ensuremath{\ensuremath{\ee.\m}}}$ rule\footnote{As an abuse of notation, if a metavariable $M$ appears somewhere
in a formula under $n$ overbars, then $M_{i_{1},\ldots,i_{n},i_{n+1},\ldots,i_{n+k}}=M_{i_{1},\ldots,i_{n}}$,
for all $n,k\geq0$. This means that in the $\rn{\vdash\ensuremath{\ensuremath{\ee.\m}}\cor\T}$
rule, as $\GG$ and $\DL$ have occurrences with zero overbars above
them, $\DL_{\t i}=\DL$ and $\GG_{\v i}=\GG$, for all $i$; thus
$\ob{\DL|\GG\vdash\ees ':\Ts '|\ob{\eps '}}$ is equivalent to $\DL|\GG\vdash\ees{'_{1}}:\Ts{'_{1}}|\ob{\eps{'_{1}}},\ldots,\DL|\GG\vdash\ees{'_{n}}:\Ts{'_{n}}|\ob{\eps{'_{n}}}$.} works in the usual way, but reports the called method as one of the
expressions effects, along with the effects of the receiver and argument
expressions. The $\rn{\vdash\textrm{\ensuremath{\ck{restrict}}}}$
rule simply reports the types and effects of $\ee$, but checks that
they are sub-effects of $\ob{\ep}$.

\subsection{Other Rules}

\vspace{0.15em}Finally, we show $\name$'s two other typing rules
that use our novel sub-effect relation:{\scriptsize{}}\\
{\scriptsize{}{\scriptsize{}$\rc{\CC\vdash\SIG\eq\ee}{\nir{\vdash\M{\SIG\eq\ee}}{\begin{array}{c}
\vdash\T\,\mmap{\ob{\X\co\CCs '}}{\ob{\Ts '}\,\x}\,\e{\ob{\ep}}\vsep\\
\ob{\X\mapsto\CCs '}|\ck{this}\mapsto\CC,\ob{\x\mapsto\Ts '}\vdash\ee:\T|\ob{\eps '}
\end{array}}{\CC\vdash\T\,\mmap{\ob{\X\co\CCs '}}{\ob{\Ts '}\,\x}\,\e{\ob{\ep}}\eq\ee}\ob{\eps '}\preceq\ob{\ep}}$\hfill{}$\nir{\vdash\SIG\eq\ee}{\begin{array}{c}
\vdash\T\,\mmap{\ob{\X\co\CCs '}}{\ob{\Ts '\,\x}}\,\e{\ob{\ep}}\vsep\\
\ob{\X\mapsto\CCs '}|\ck{this}\mapsto\CC,\ob{\x\mapsto\Ts '}\vdash\ee:\T|\ob{\eps '}
\end{array}}{\CC\vdash\T\,\mmap{\ob{\X\co\CCs '}}{\ob{\Ts '\,\x}}\,\e{\ob{\ep}}\eq\ee}\ob{\eps '}\preceq\ob{\ep}$\hfill{}\hbox{}}{\scriptsize\par}

{\scriptsize{}\hrule}{\scriptsize\par}

\noindent {\scriptsize{}\hbox{}$\rc{\vdash\ob{\SIG\eq\ee}\vartriangleleft\SIGs '}{\nir{\ob{\SIG\eq\ee}\vartriangleleft\SIGs '}{\begin{array}{c}
(\Ts{_{1}}\,\mmap{\ob{\Xs{_{2}}\co\CCs{_{1}}}}{\ob{\Ts{'_{1}\,\xs{_{1}}}}}\,\e{\ob{\eps{_{1}}}}\_)[\ob{\Xs{_{2}}\coloneqq\Xs{_{1}}}]\in\ob{\SIG\eq\ee}\vsep\\
\DL=\ob{\Xs{_{2}}\mapsto\CCs{_{2}}}\hsep\DL\vdash\ob{\eps{_{1}}}\preceq\ob{\eps{_{2}}}\vsep\\
\DL\vdash\Ts{_{1}}\leq\Ts{_{2}}\hsep\ob{\DL\vdash\Ts{'_{2}}\leq\Ts{'_{1}}}\hsep\ob{\DL\vdash\CCs{_{2}}\leq\CCs{_{1}}}\hsep
\end{array}}{\vdash\ob{\SIG\eq\ee}\vartriangleleft\Ts{_{2}}\,\mmap{\ob{\Xs{_{2}}\co\CCs{_{2}}}}{\ob{\Ts{'_{2}\,\xs{_{2}}}}}\,\e{\ob{\eps{_{2}}}}}}$\hfill{}$\mathclap{\nir{\ob{\SIG\eq\ee}\vartriangleleft\SIGs '}{\begin{array}{c}
(\Ts{_{1}}\,\mmap{\ob{\Xs{_{2}}\co\CCs{_{1}}}}{\ob{\Ts{'_{1}\,\xs{_{1}}}}}\,\e{\ob{\eps{_{1}}}}\_)[\ob{\Xs{_{2}}\coloneqq\Xs{_{1}}}]\in\ob{\SIG\eq\ee}\vsep\\
\DL=\ob{\Xs{_{2}}\mapsto\CCs{_{2}}}\hsep\DL\vdash\ob{\eps{_{1}}}\preceq\ob{\eps{_{2}}}\vsep\\
\DL\vdash\Ts{_{1}}\leq\Ts{_{2}}\hsep\ob{\DL\vdash\Ts{'_{2}}\leq\Ts{'_{1}}}\hsep\ob{\DL\vdash\CCs{_{2}}\leq\CCs{_{1}}}\hsep
\end{array}}{\vdash\ob{\SIG\eq\ee}\vartriangleleft\Ts{_{2}}\,\mmap{\ob{\Xs{_{2}}\co\CCs{_{2}}}}{\ob{\Ts{'_{2}\,\xs{_{2}}}}}\,\e{\ob{\eps{_{2}}}}}}$\hfill{}\hbox{}}{\scriptsize\par}}{\scriptsize\par}

\noindent We use judgements of the form $\CC\vdash\SIG\eq\ee$ to
check that each method, $\SIG\eq\ee$, of every class, $\CC$, is
well typed. The $\rn{\vdash\SIG\eq\ee}$ rule checks that $\SIG$
is well-typed, and that the expression $\ee$ is well-typed under
the appropriate variable and type-parameter contexts. The rule also
checks that the effects of $\ee$ are sub-effects of those in the
$\ck{effect}$ annotation of $\SIG$.

To check that a class with methods $\ob{\SIG\eq\ee}$ correctly implements
interfaces, we use judgements of the form $\vdash\ob{\SIG\eq\ee}\vartriangleleft\SIGs '$,
for each method signature $\SIGs '$ in each implemented interface.
Our $\rn{\ob{\SIG\eq\ee}\vartriangleleft\SIGs '}$ rule allows the
usual refinements: more specific return types, more general argument
types and type-parameter bounds, as well as different (type) parameter
names. In addition, we allow refining the $\ck{effect}$ annotation:
a method's declared effects must be sub-effects of those declared
by the signature that is being implemented.

The rest of our type system is otherwise standard, and is presented
in Appendix \ref{sec:Auxiliary}. We also assume the usual rules that
classes, methods, fields, type-parameters and method-parameters are
not declared with duplicate names.

\section{Extensions}\label{sec:Extensions}

The discussion and formalism presented above for $\name$ is missing
many potentially useful features; here we present an outline for three
such extensions that would be worth exploring further in future work.
In particular, we discuss allowing useful and sound dynamic code loading
and reflective method invocation; adding support for both sub-type
and sub-effect variance, with respect to generic type parameters;
and the ability to alter the effects of a method based on the context
it is called in. As these extensions are still works in progress,
we do not formally define their semantics, but rather given an informal
explanation for their semantics and use case.

\subsection{Reflection and Dynamic Class Loading}

One common feature in many OO languages is reflection, which can be
used to \emph{dynamically} invoke a method, without statically knowing
weather it exists or has an appropriate signature. For example, consider:

\begin{lstlisting}
	#reflect-invoke(Console.#read(), 1)
\end{lstlisting}

\noindent Here $\cf{reflect-invoke}$ will, at runtime, find a method
with whatever name the user typed into the console, check that it
accepts an argument/receiver of type $\ct{Int}$, and then invoke
it with $\cl 1$. But what should the \emph{effects} of $\cf{reflect-invoke}$
be? As we statically don't know anything about what method it may
try and invoke, the only safe option would be for $\cf{reflect-invoke}$
to be declared with $\e{\*}$. Alternatively, instead of $\cf{reflect-invoke}$
being a library function, we could make it a primitive form of expression,
which takes an effect as an argument: $\ck{invoke[}\ob{\ep}\ck{](}\ee,\ob{\ees '}\ck )$.
This expression will have have the effects $\ob{\ep}$, as well as
the effects of $\ee$ and $\ob{\ees '}$. This will behave like $\cf{reflect-invoke(}\ee,\ob{\ees '}\cf )$,
except that it will dynamically check that the called method is a
sub-effect of $\ob{\ep}$.

But what about dynamic class loading? This is one common use case
for reflection, where a method won't actually exist until the program
is run. Consider for example the following class:

\begin{lstlisting}
class C { static Void #foo() effect[C.#foo()] = ...; }
\end{lstlisting}

\noindent Suppose we want to dynamically load and invoke $\c{\t C.\f{foo}}$:

\begin{lstlisting}[escapechar={`}]
static Void #bar() effect[*] = (
	#load-class(`$\cM{/* code for \t{C} above */\f{)}}$`; 
	invoke[](`$\cl{"}$`C.#foo`$\cl{"}\ck{)}\cb{)}$`;
\end{lstlisting}

Will $\cf{bar()}$ produce a runtime error? $\c{\t C.\f{foo}}$ is
declared as effectful, so the dynamic invocation of it (with the empty
effect) will not work. However, the reason it fails is because $\c{\t C.\f{foo}}$
has $\c{\t C.\f{foo}}$ in its $\ck{effect}$ annotation; since $\c{\t C.\f{foo}}$
did not exist when the $\ck{invoke}$ expression was typed/compiled,
the only way to to call such dynamically loaded methods would be with
the $\*$ effect. Instead, we propose ignoring the presence of any
methods that did not exist when the $\ck{invoke}$ expression was
type checked; formally this means that $\ck{invoke[}\ob{\ep}\ck{](}\cl "\t T.\m\cl ",\ldots\ck )$
will check that $\t T.\m\preceq_{\SI_{\t 0}}\ob{\ep}$ where $\SI_{\t 0}$
is the class table when the expression was type checked, and $\preceq_{\SI_{\t 0}}$is
like $\preceq$, but also with the following variation of the $\rname$
rule:

{\scriptsize{}\hbox{}\hfill{}$\nir{\ensuremath{\preceq_{\mt{\SI_{_{0}}}}}}{\begin{array}{c}
\SI(\DL(\T)\,.\m)=\_\,\mmap{\_}{\_}\,\e{\ob{\eps '}}\,\_\vsep\\
\ob{\eps '}\mathrel{\preceq_{\mt{\SI_{_{0}}}}}\Ts '.\ms ',\ob{\ep}\hsep\Ts '.\ms '\notin\op{dom}(\SI_{\t 0})
\end{array}}{\DL\vdash\Ts '.\ms '\preceq_{\mt{\SI_{_{0}}}}\ob{\ep}}$\hfill{}\hbox{}}{\scriptsize\par}

\noindent A similar rule could also be added to allow calling methods
with private methods in their effect annotation: in which $\SI_{\t 0}$
will only contain declarations for \emph{accessible }methods. 

Though it is clear to us how $\ck{invoke}$ should work, formalising
it would require non-trivial modifications to the FGJ runtime semantics.

\subsection{Generic Type Parameter Variance}\label{subsec:Variance}

One likely useful addition is support for declaration-site\footnote{This is different to the more expressive, complex, and verbose \emph{use-site
}variance offered by Java's wildcards; although we could also extend
$\name$ to support such a feature.} generic type parameter variance, as is supported by OO languages
such as Scala, C\textsuperscript{\ensuremath{\sharp}}, and Kotlin.
For example, consider the following interface for iterators:

\begin{lstlisting}
interface Iterator<out T> { T #next(); }
\end{lstlisting}
Here, as in C\textsuperscript{\ensuremath{\sharp}} and Kotlin, we
use $\ck{out}$ $\ct T$ to declare that $\ct T$ is a \emph{covariant}
parameter: for any types $\t T$ and $\Ts '$, if $\T$ is a subtype
of $\Ts '$ then $\ct{Iterator\a{\T}}$is also a subtype of $\ct{Iterator\a{\Ts '}}$.
This makes sense since one can use an $\ct{Iterator\a{\ct{String}}}$
as if it were an $\ct{Iterator\a{\ct{Object}}}$, since calling $\cf{next}$
on the former produces a $\cf{String}$, which is a sub-type of $\cf{Object}$
(the type returned by calling $\cf{next}$ on the later). The usual
rule for when an $\ck{out}$ $\ct T$ declaration is valid is that
within the method signatures of the class\emph{,} $\ct T$ only appears
in a covariant position: in particular, the return type of a method
is a covariant position.

It would make sense to use this feature together with effect annotations
as well, for example consider the following contrived interface for
logging objects:

\begin{lstlisting}
interface Logger<in T> {
	Void #log(T x) effect[File-System.#append];
	Void #log-all(Iterator<T> x) effect[Logger<T>.#log]; }
\end{lstlisting}

\noindent Here we use $\ck{in}$ $\ct T$ to declare that $\ct T$
is a \emph{contravariant} parameter: for any types $\t T$ and $\Ts '$,
if $\T$ is a subtype of $\Ts '$ then $\ct{Logger\a{\T}}$is a \emph{supertype}
of $\ct{Logger\a{\Ts '}}$. This should be allowed when $\T$ only
appears in contravariant positions. In particular, an argument type
of a method is a contravariant position. We suggest also making the
$\ck{effect}$ annotation a\emph{ covariant} position: if $\t T$
is a subtype of $\Ts '$, then our typing rules enforce that $\T.\m$
is a sub-effect of $\Ts '.\m$. When types are used as generic parameters
however, co and contra-variance gets more complicated. In general
if $\CCs{\a{\T}}$ is in a covariant position, and $\CC$ is declared
as taking a covariant type parameter, then $\t T$ is also in a covariant
position, otherwise if it is a contravariant parameter, then $\T$
is in a contra-variant position. On the other hand, if $\CCs{\a{\T}}$
is in a contravariant position, then $\T$ is in a covariant (or contravariant)
position if $\CC$ takes a contravariant (or, respectively, covariant)
type parameter. The above code is sound since $\ct T$ only appears
in contra-variant positions: the argument type of $\cf{log}$, as
a covariant parameter to $\ct{Iterator}$ in the argument type of
$\cf{log-all}$, and as a contravariant parameter to $\ct{Logger}$,
which appears in the $\ck{effect}$ annotation (a covariant position)
of $\cf{log-all}$.

Now consider generic method parameters:

\begin{lstlisting}
static Void #secure-hash<out X: Hashable>(X x)
	effect[#secure-hash, X.#hash] = ...;
\end{lstlisting}

\noindent Since the only use of $\c{\ct X}$ in the effect annotation
is in a covariant position, this example should type-check and mean
that the \emph{effect} of $\cf{secure-hash}$ is covariant with respect
to $\ct X$: if $\T$ is a sub-type of $\Ts '$ then $\cf{secure-hash\a{\T}}$
is a sub-effect of $\cf{secure-hash\a{\Ts '}}$. This relation makes
intuitive sense: if $\T$ is a subtype of $\Ts '$ then $\T.\cf{hash}$
is no more effectful than $\Ts '.\cf{hash}$, and so $\cf{secure-hash\a{\T}}$
cannot be more effectful than the effect declaration of $\cf{secure-hash\a{\Ts '}}$.
Note that at runtime however, calling $\cf{secure-hash\a{\T}}$ could
perform more effectful behaviour than calling $\cf{secure-hash\a{\Ts '}}$,
but no more than the effect declaration of $\Ts '.\cf{hash}$ allows.

We have not formalised this kind of variance since this would be merely
a simple straightforward extension over non-trivial prior work \cite{Variance_and_Generalized_Constraints_for_C|sharp_Generics,Variant_parametric_types:_A_flexible_subtyping_scheme_for_generics}.

\subsection{Effect Redirection}

One potentially useful feature would be to add something similar to
effect \emph{handlers} to $\name$, as it could be particularly useful
to redirect an effectful method to a less effectful one:

\begin{lstlisting}
static Void #untrusted() effect[Console.#print] = ...;
static Void #fake-print(String s) effect[] = skip;
static Void #sandbox() effect[] =
	redirect[Console.#print = #fake-print] #untrusted();
\end{lstlisting}
The idea is that $\ck{redirect[}\T.\m\eq\Ts '.\ms '\ck ]$ $\ee$
would execute $\ee$, but at runtime, whenever it tries to call $\T.\ms{\p{\ob{\ee}}}$,
it actually calls $\Ts '.\ms{'\p{\ob{\ee}}}$.\footnote{We leave it to future work to determine how this should behave when
either $\T.\m$ or $\Ts '.\ms '$ are instance methods. For example,
if $\T.\m$ is an instance method, it could redirect calls of the
form $\ee.\ms{\p{\ob{\ees '}}}$ to $\Ts '.\ms{'\p{\ee\c ,\ob{\ees '}}}$.
We could also allow $\T.\m\eq\ee.\ms '$ to redirect calls of the
form $\T.\ms{\p{\ob{\ees '}}}$ to $\ee.\ms{'\p{\ob{\ees '}}}$; this
would be particularly useful if $\ee$ were a lambda expression.} Thus in the above example, we can can be sure that $\cf{sandbox}$
prints nothing to the console. However formalising this with our effect
system is non-trivial, for example suppose $\cf{untrusted}$ was instead
declared with:

\begin{lstlisting}
	effect[Console.#print, Console.#read]
\end{lstlisting}

\noindent What should the effects of the $\ck{redirect}$ expression
be? We could choose an upper bound such as $\ct{Console}\c .\cf{read}$,
but that throws away the information that it was called through $\cf{untrusted}$.
Instead we are thinking of adding `substitution effects' of the
form $\ep\cf [\T.\m\eq\ob{\eps '}\cf ]$, meaning the effect $\ep$,
except that any calls to $\T.\m$ will actually have sub-effects of
$\ob{\eps '}$. In particular, we would have that $\T.\m\cf [\T.\m\eq\ob{\ep}\cf ]\preceq\ob{\ep}$.
With this, the effects of\vspace{0.1\baselineskip} $\ck{redirect[}\T.\m\eq\Ts '.\ms '\ck ]$
$\ee$ would be $\ob{\ep\cf [\T.\m\eq\Ts '.\ms '\cf ]}$ (where $\ee$
has effects $\ob{\ep}$). There are other sub-effect rules we could
\vspace{0.1\baselineskip} include: if $\ep\preceq\ob{\eps '}$, then
$\ep\cf [\T.\m\eq\ob{\eps{''}}\cf ]\preceq\ob{\eps '\cf [\T.\m\eq\ob{\eps{''}}\cf ]}$;
if $\ob{\eps '}\preceq\ob{\eps{''}}$, then $\ep\cf [\T.\m\eq\ob{\eps '}\cf ]\preceq\ep\cf [\T.\m\eq\ob{\eps{''}}\cf ]$;
and $\*\preceq\*\cf [\T.\m\eq\ob{\ep}\cf ]$. However we are not sure
if there are more rules that would be useful, and if there is a smaller
but equivalent set. Though our example shows that the type system
using such substitution effects internally could be useful, we are
not sure if exposing such additional complexity would be worthwhile.
For example, would a programmer file the following annotation on $\cf{sandbox}$
useful or unnecessarily complex:

\begin{lstlisting}
static Void #sandbox()
	effect[#untrusted[Console.#print = #fake-print]]
\end{lstlisting}

\section{Larger Examples}\label{sec:Example}

In this section we show two larger examples demonstrating the practical
benefits of $\name$'s effect system. First we show how $\name$ can
help prevent security vulnerabilities by restricting how code interacts
with a database. Then we show how a graphical program can contain
code for untrusted advertisements, whilst being sure they will not
interfere with the rest of the program's GUI.

\subsection{Controlling Access to a Database}

Consider an interactive program that wishes to communicate with an
SQL database. There are several security vulnerabilities that could
arise here: user input could be mistakenly processed as an SQL query
(i.e. SQL injection), arbitrary SQL queries could be executed by untrusted
components of the program, and sensitive data resulting from queries
could be handled by such untrusted components. Here we show how our
effect system can help prevent such potential vulnerabilities.

Consider the following sketch of a class for prepared statements:

\begin{lstlisting}[escapechar={`}]
class SQL-Statement {
	// Make a new prepared statement, with `$\cl{"?"}$` as placeholders
	// for arguments (here $Socket.#read-write is a foreign
	// method that communicates with the database)
	static SQL-Statement #prepare(String s)
		effect[#prepare, Socket.#read-write] = ...;

	// Execute the query replacing any `$\cl{"?"}$`s with arguments
	SQL-Result #execute(List<String> arguments)
		effect[#execute, Socket.#read-write]  = ...;

	// Assuming a private constructor
	... }
\end{lstlisting}

The idea is that a $\ct{Statment}$ is first made by calling $\cf{prepare}$,
using place-holders (`?') wherever values from user-input might
be inserted, and then providing the values for such place-holders
by calling $\cf{execute}$. For example, we could safely obtain the
age of a given user:

\begin{lstlisting}[escapechar={`},mathescape=true]
SQL-Statement.#prepare(`$\cl{"}$`SELECT age FROM Users WHERE name = `$\cl{?"}$`$)
	.#execute({Console.#read-line()})
\end{lstlisting}

This code is resilient to SQL injection because the arguments parsed
to $\cf{execute}$ will not be treated as SQL syntax, but rather as
plain text with no special meaning for any characters. This means
that no matter what input the user writes, $\cf{execute}$ will return
the $\cv{age}$ of all $\ct{Users}$ with the given $\cv{name}$ (if
any). SQL injection can still occur if this API is misused:

\begin{lstlisting}[escapechar={`},mathescape=true]
SQL-Statement.#prepare(`$\cl{"}$`SELECT age FROM Users WHERE name = `$\cl{\textquotesingle"}$`$
	+ Console.#read-line() + "'"`$\cf{)}$`.#execute({})
\end{lstlisting}

Now if the user types in `$\cl{\textquotesingle}\c ;$ $\ck{DROP}$
$\ck{TABLE}$ $\ct{Users}$ $\cM{--}$', the $\ct{Users}$ table
will be deleted!

Because $\cf{prepare}$ is declared as effectful, and is not a sub-effect
of $\cf{execute}$, we only need to consider the code of any methods
declared with super-effects of $\cf{prepare}$, and check that no
such code incorrectly passes user input to $\cf{prepare}$. Any other
code, even if it calls the $\cf{execute}$ method, cannot possibly
cause SQL injection:

\begin{lstlisting}[escapechar={`}]
// Might cause SQL injection (perhaps @param contains user
// input)
List<SQL-Statement> #setup-queries(Object param)
	effect[SQL-Statement.#prepare] = ...;


// Can't cause SQL injection, but can execute @queries
SQL-Result #run-queries(List<SQL-Statement> queries)
	effect[SQL-Statement.#execute, Console.#read] = ...;
\end{lstlisting}

We would also like to prevent untrusted code from looking at the potentially
sensitive information retrieved by such queries. We can restrict this
by making the results of such queries be wrappers over strings:

\begin{lstlisting}[escapechar={`}]
class SQL-Result(private String result) {
	String #value() effect[#value] = result; }
\end{lstlisting}

Because $\cf{value}$ is effectful, code without such an effect will
see an $\ct{SQL-Result}$ as a black box: such code can pass around
$\ct{SQL-Result}$s, but cannot inspect their internal state. Thus
in order to understand how results from database queries are used,
we do not need to reason about how $\ct{SQL-Results}$ are passed
around, we only need to look at methods declared with (super effects
of) the $\cf{value}$ effect and check that they do not inappropriately
process or pass the result of $\cf{value}$:

\begin{lstlisting}[escapechar={`}]
Bool #is-adult(SQL-Result age) effect[SQL-Result.#value] =
	Int.#parse(age.#value) `$\mathtt{\geq}$` 18`{}`;

// Can't know the exact age of users, but can check if they
// are adults
Void #do-stuff(SQL-Statement age-query) 
	effect[SQL-Statement.#execute, #is-adult] = ...;
\end{lstlisting}

\subsection{Controlling Access to GUI Widgets}

Suppose we have an app that contains untrusted advertisements which
need to maintain their own GUI widgets, but must not modify other
parts of the app's GUI. One way of doing this is by ensuring that
advertisement code is only given references to widgets it is allowed
to modify \cite{Reasoning_about_Object_Capabilities_with_Logical_Relations_and_Effect_Parametricity}.
This approach works best when there is a common ancestor widget all
of whose descendants should be modifiable by the ad. Suppose we instead
have a complicated GUI with many nested widgets, where some nodes
should be modifiable by the advertisement, but not their parents or
children.

We can use our effect system to statically ensure that advertisements
do not attempt to modify properties of widgets they are not allowed
to, and we can give the advertisement a reference to the top-level
widget, even if it is not itself modifiable by the advertisement.
We can encode such a permission using an abstract method of an interface:

\begin{lstlisting}
interface Permission { Void #modify() effect[#modify]; }
\end{lstlisting}

This interface is not designed to be instantiated by any concrete
classes (although doing so will not cause problems), rather its method
will be used in effect annotations. The idea is that a method marked
with $\e{\ct{Permission}\c .\cf{modify}}$ may modify properties of
a widget, and can only be (directly) called by methods mentioning
it or $\ct{\ct{Permission}}\c .\cf{modify}$ in their effect annotation.

We can create a proper sub-effect of $\ct{\ct{Permission}}\c .\cf{modify}$
by creating a sub-interface of $\ct{Permission}$:

\begin{lstlisting}
interface Ad-Permission: Permission { }
\end{lstlisting}

The $\ct{Ad-Permission}$ interface will inherit the method of $\ct{Permission}$
along with its effect annotation: $\ct{Ad-Permission}$ contains a
$\cf{modify}$ method with `$\ct{\ct{Permission}}\c .\cf{modify}$'
as its effect annotation. Since effects are covariant with respect
to their receiver, we will have that $\ct{Ad-Permission}\c .\cf{modify}$
is a sub-effect of $\ct{\ct{Permission}}$$\c .\cf{modify}$; however
the converse does not hold. Thus the $\ct{\ct{Permission}}\c .\cf{modify}$
effect is sufficient to call methods with $\ct{Ad-Permission}\c .\cf{modify}$
in their effect annotation, but not vice versa.

At first glance it may seem like widget \emph{classes} will need to
be specifically written for the use of ads by mentioning the $\ct{Ad-Permission}$
interface. We can make such classes generic, thus encoding permission-generacity
\cite{Generic_ownership_for_generic_Java}, and let \emph{object}
creation determine what permission to require:

\begin{lstlisting}
interface Widget { ...`{}` }
// A sidebar with a stack of widgets on top of eachother
class Sidebar<P: Permission>(List<Widget> c): Widget {
	// Gets the @index`\textsuperscript{$\cM{th}$}` widget
	Widget #get(Nat index) effect[] = c.#get(index);

	// Sets the @index`\textsuperscript{$\cM{th}$}` widget
	Void #set(Nat index, Widget widget) effect[P.#modify] =
		c.#set(index, widget); ... }
\end{lstlisting}

The idea is that the $\ct P\c .\cf{modify}$ effect is sufficient
to modify the children of a $\ct{Sidebar\a{\ct P}}$.\footnote{Since the $\ct{Sidebar}$ class only uses $\ct P$ in effect annotations,
we could have declared it with $\ck{out}$, thus marking it is as
covariant (see \subsecref{Variance}). With this, one could use a
$\ct{Sidebar\a{\ct{Ad-Permission}}}$ as if it were a $\ct{Sidebar\a{\ct{Permission}}}$.} We can now safely give an arbitrary widget to an advertisement, and
be sure that it won't modify any sidebars that were not created with
$\ct{Ad-Permisson}$:

\begin{lstlisting}
interface Ad {
	Void #update(Sidebar<Permission> sb)
		effect[Ad-Permission.modify]; }

class Example-Ad(): Ad {
	Void #update(Sidebar<Permission> sb) 
		effect[Ad-Permission.modify] = (
			`$\cM{// Cast}$` sb`$\cM{.}$`#get(0) `$\cM{to a type we can modify}$`
			(`$\ct{(}$`Sidebar<Ad-Permission>`$\ct{)}$`(sb.#get(0))).#set(...)`$\cb{))}$`;}
\end{lstlisting}

\begin{lstlisting}
// Some nested sidebars, those created with $<$Ad-Permission> 
// are modifiable by $Ad, but those with $<$Permission> are not.
var sidebar1 = new Sidebar<Permission>(...);
var sidebar2 = new Sidebar<Ad-Permission>({sidebar1, ...});
var sidebar3 = new Sidebar<Permission>({sidebar2, ...});

Ad ad = ...;
// Regardless of what the dynamic type of @ad is, this won't 
// modify @sidebar1 or @sidebar3, but could modify  @sidebar2.
ad.#update(@sidebar3);
\end{lstlisting}

For the above dynamic casts in $\ct{Example-Ad}$ to be sound we need
generic reification (as in C\textsuperscript{\ensuremath{\sharp}})
instead of type erasure (as in Java). Though this does not require
reification of any \emph{effect }information (such as the $\ck{effect}$
annotations of methods or the sub-effect relation).

\section{Soundness}\label{sec:Soundness}

In order to ensure that the reasoning we showed in \secref[s]{Reasoning}
\& \ref{sec:Example} is correct, we need to be sure that our type
system enforces our interpretation of effects. In particular, we need
to know what the \emph{direct} effects of a method call could be,
and what the \emph{indirect} effects could be. To do this we will
informally sketch the proofs of the two soundness statements.

First we define \emph{direct soundness}: the effect annotations of
method declarations are respected, even in the presence of dynamic
dispatch and sub-typing:

\vspace{1.5\smallskipamount}

\textbf{Direct Soundness:} \emph{If a method $\CC.\mms{\a{\ob{\T}}}$
is annotated with $\e{\ob{\ep}}$, then a well-typed call to $\mms{\a{\ob{\T}}}$
on a receiver of (static) type $\CC$ will directly reduce to an expression
whose effects are (sub-effects of) $\ob{\ep}$.}

\emph{Proof Sketch}: Our typing rule for method bodies, $\rn{\vdash\SIG\eq\ee}$,
ensures that the effects of the method body (before any substitutions
have occurred) are sub-effects of its $\ck{effect}$ annotation, which
our refinement rule, $\rn{\ob{\SIG\eq\ee}\vartriangleleft\SIGs '}$,
ensures is a sub-effect of that of $\CC.\mm$.

Because our $\rn{\preceq\T}$ rule ensures that method effects are
covariant with respect to their receiver, and type preservation (which
we inherit from FGJ) ensures that evaluation can only refine types,
we can be sure that any (type) parameter substitutions that are performed
by the method call can only refine the effects of the method body.
Thus, after substitution has been performed, the effects of the method
body will be sub-effects of those before substitution, and hence will
also be sub-effects of the $\ck{effect}$ annotation of the (substituted)
method declaration for $\CC.\mms{\a{\ob{\T}}}$. $\QED$

\vspace{1.5\smallskipamount}

Secondly, we define \emph{indirect soundness}: the \emph{only }effectful
behaviour an expression with (sub-effects of) $\ob{\ep}$ can perform
is that allowed by the methods listed in $\ob{\ep}$:

\vspace{1.5\smallskipamount}

\textbf{Indirect Soundness: }\emph{If a well-typed expression has
sub-effects of $\ob{\ep}$, and after reducing it any number of times,
it contains a call to $\CC.\m$, then either: (a) $\CC.\m$ (or a
method it overrides) is in $\ob{\ep}$, (b) the $\ck{effect}$ annotation
of $\CC.\m$ is (a sub-effect of) $\ob{\ep}$, or (c) the call to
$\CC.\m$ was (indirectly) introduced by the reduction of a call to
a method (overriding one) in $\ob{\ep}$.}\footnote{Case (\emph{c}) can be thought of as saying the call to $\CC.\m$
is below the stack frame of a method (overriding one) one in $\ob{\ep}$.}

\emph{Proof Sketch}: This can be proved by induction on the depth
of the call stack. In the base case the call was present in the initial
expression. Because the receiver of the call may have been (partially)
reduced, the call to $\CC.\m$ was originally a call to $\CCs '.\m$,
for some $\CCs '\geq\CC$. Thus, by transitivity and covariance of
our sub-effect relation (the $\rn{\preceq Trans}$ and $\rn{\preceq\T}$
rules), we have $\CC.\m\preceq\CCs '.\m\preceq\ob{\ep}$. By analysing
our sub-effect rules, it can be seen that either $\CC.\m$ (or a method
it overrides) must be in $\ob{\ep}$ (because the $\rn{\preceq\T}$,
$\rn{\preceq Ref}$, $\rn{\preceq Comp}$, and $\rn{\preceq\ensuremath{\emp}}$
rules apply), or the $\rname$ rule applies (this trivially holds
if $\*$ is in $\ob{\ep}$ or $\CC.\m$ is uneffectful). Either way,
case (\emph{a}) or (\emph{b}) holds.

In the inductive case, we will have that the call to $\CC.\m$ was
directly\emph{ }introduced by a call to a method $\CCs '.\ms '$ satisfying
case (\emph{a}), (\emph{b}), or (\emph{c}). If $\CCs '.\ms '$ satisfied
case (\emph{a})\emph{ }or (\emph{c}), then clearly $\CC.\m$ now satisfies
case (\emph{c}). If $\CC.\m$ does not satisfy case (\emph{c}), then
we must have that $\CCs '.\ms '$ satisfied case (\emph{b}). Let $\ob{\eps '}$
be the effect annotation of $\CCs '.\ms '$, then by direct soundness,
case (\emph{b}), and our sub-effect transitivity rule, $\rn{\preceq Trans}$,
we have that $\CC.\m\preceq\ob{\eps '}\preceq\ob{\ep}$. By the same
logic we used above for the base case, we can conclude that case (\emph{a})
or (\emph{b}) holds for $\CC.\m$.$\QED$

\vspace{1.5\smallskipamount}

Note that case (\emph{c}) only says the call to $\CC.\m$ was introduce
by a call to one in $\ob{\ep}$. To further reason about $\CC.\m$,
one can either analyse the source code of the methods in $\ob{\ep}$,
or use direct and indirect soundness on their effect annotations.

Together \emph{direct soundness }and \emph{indirect soundness} ensure
our key property: the only \emph{effectful} behaviour a method $\T.\m$
declared with $\ck{\e{\ob{\ep}}}$ can perform is that of each $\ob{\ep}$.
Our soundness statements also allow us to reason as to what methods
can never be called: if another method $\Ts '.\ms '$ is not a sub-effect
of $\ob{\ep}$, nor is it a sub-effect of the effects of any methods
in $\ob{\ep}$, and so on transitively, we can be sure that calling
$\T.\m$ will not (indirectly) call $\Ts '.\ms '$. We can also perform
such reasoning on $\ck{restrict}$ expressions, since their typing
rule, $\rn{\vdash\textrm{\ensuremath{\ck{restrict}}}}$, enforces
a bound on the effect of their bodies.

\section{Related Work}\label{sec:Related}

There are three main techniques in the literature for reasoning over
effectful operations: type and effect systems, monads, and object-capabilities.
The former two are most common in functional programming languages,
whereas the latter is naturally used in OO languages.

\subsection{Type and Effect Systems}

Most effect type systems in the literature feature a specific set
of effects designed for a particular purpose \cite{Type_and_Effect_Systems}.
The application of Hindley--Milner style polymorphism and type inference
\cite{Convenient_Explicit_Effects_using_Type_Inference_with_Subeffects,Koka:_Programming_with_Row-polymorphic_Effect_Types,Lightweight_Polymorphic_Effects}
to such effect systems has been studied heavily. Though most effect
systems have focused on functional programming languages, research
has also investigated object-oriented languages, such as for traditional
region based effects \cite{An_Object-Oriented_Effects_System}, and
controlling access to UI objects \cite{JavaUI:_Effects_for_Controlling_UI_Object_Access}.

An effect system for control flow analysis \cite{Type_and_Effect_Systems,Separate_Abstract_Interpretation_for_Control-Flow_Analysis,Effect_Systems_with_Subtyping}
annotates lambda expressions with `labels' and then uses such labels
as effects. This system is not intended to be used by programmers
directly, rather it would be used internally by automated analyses,
as it requires lambdas to list the label of \emph{every} other lambda
they may (indirectly) call. Its sub-effect relation is much simpler
than ours, allowing this system to be presented with a sound and complete
algorithm that will take an expression with no effect annotations,
and produce a minimally annotated version.

A `generic' framework for effect systems has also been presented
with a language definition parametrised by `effect disciplines'
\cite{A_Generic_Type-and-Effect_System}. An `effect discipline'
defines what constitutes an effect, how the effects of each expression
form are computed, and when an expression is valid for a given set
of effects. This system is more general than ours, as effect disciplines
may allow an uneffectful expression to contain effectful sub-expressions,
e.g. a $\ck{throw}$ expression inside a $\ck{try}$--$\ck{catch}$
expression, where the $\ck{throw}$ itself has a `throw' effect
but the enclosing $\ck{try}$--$\ck{catch}$ does not.

A distinct, but related concept is that of effect \emph{handler} systems
\cite{Do_Be_Do_Be_Do,Handlers_of_Algebraic_Effects}, which were originally
presented for pure functional languages, but have also been extend
to object-oriented languages \cite{JEff:_Objects_for_Effect}. These
are essentially a system for checked and resumable exceptions; however
the main motivation for their use is also to reason over side-effects.
For example, a method $\cf{foo}$ could be declared as raising an
effect/exception `$\cf{read}$', the caller of this method can then
provide a handler for $\cf{read}$, so that when $\cf{foo}$ raises
the $\cf{read}$ `exception', a value is returned like an ordinary
method call; alternatively the handler could behave like a normal
exception and terminate execution of $\cf{foo}$. The main advantages
of this approach is that one can determine the effectful operations
a method may perform by simply looking at its signature, as well as
allowing these operations to be arbitrarily redefined. An effect handler
system that allows effects to declare default handlers (in case the
effect were leaked from the main function) could be seen as an alternative
to our system; however they require that \emph{every} effect a method
leaks to be explicitly listed in their signature, as well as requiring
explicit effect handling to allow reasoning like that in our $\cf{log}$
example from \subsecref{Reasoning-Indirect}.

\subsection{Monads}

In functional programming (such as Haskell) the most common tool for
side effect reasoning is that of monads; however one usually has to
write code in a specific `monadic' style that is very alien compared
to typical imperative code, for example instead of writing $\cf{print(readLn())}$,
in Haskell, one would write $\c{\cf{readLn}\ >>=\ \cf{print}}$. Or
using the less flexible $\ck{do}$ syntax sugar, $\ck{do}\ \ck{\{}\cv{line}\ \c{<-}\ \text{\ensuremath{\cf{readLn}}}\c ;\ \cf{print}\ \cv{line}\k{\}}$.
However, some monads can even be `unwrapped', for example Haskell's
$\cf{runST}$ function allows stateful computations of type $\forall\t s\c .\ct{ST}\,\t s\,\t a$
(which cannot access externally accessible memory) to be executed
and return a pure value (of type $\t a$).

Prior work has also shown how effect systems can be represented as
monads \cite{Parametric_Effect_Monads_and_Semantics_of_Effect_Systems,The_Marriage_of_Effects_and_Monads},
for example the $\ct{effect-monad}$ package \cite{Embedding_effect_systems_in_Haskell}
uses advanced Haskell type system extensions provided by GHC to allow
easily encoding user-defined effects as monads. In particular, it
supports sub-effecting in ways similar to $\name$. This library is
more general than $\name$, but is more verbose: both in the use and
definition of effects.

\subsection{Object-Capabilities}

Object-capabilities appear to be the most active research area in
OO languages for side-effect reasoning \cite{Capability_Myths_Demolished,Declarative_Policies_for_Capability_Control,Reasoning_about_Object_Capabilities_with_Logical_Relations_and_Effect_Parametricity}.
The object-capability model relies on two language guarantees: references
cannot be forged and calling a method on an object is the only way
to perform restricted operations. For example, in this model one could
have a $\ct{File}$ class, where calling a $\cf{write}$ method on
an instance of this class is the only way to (directly) $\cf{write}$
to a file (compare this with $\name$, in which the $\cf{write}$
method can be a static method). A $\ct{File}$ object is a `capability'
object, and can only be obtained by making a new one (from another
pre-existing capability object, such as a $\ct{Directory}$ one),
by receiving it as an argument to a method, or through a field of
a reachable object. Typically, the main function of the program will
be initially given an, all-powerful capability from which other capability
objects can be made (compare this with $\name$'s $\*$ effect). Compared
to effect systems, these simple requirements do not require type system
features or annotations, allowing for example Joe-E \cite{Joe-E:_A_Security-Oriented_Subset_of_Java},
a subset of Java, to be created by simply disallowing problematic
features. In particular it heavily restricts global variables and
static methods in order to prevent `ambient authority', i.e., the
ability to perform restricted operations without being explicitly
passed a capability object. $\name$ does not enforce the object-capability
style, so it does not need such heavy restrictions.

A key advantage of object-capability based systems is that new capability
objects can be \emph{dynamically} created. For example one can create
a $\ct{File}$ object and pass it to a method, granting that method
the capability to read only a single (dynamically chosen) file. In
contrast, our effect system only allows \emph{statically} creating
new effects (i.e. methods), such as our $\cf{log}$ method from \subsecref{Reasoning-Indirect},
which can only write to the statically chosen file. However, $\name$
allows a less restrictive form of the object-capability style as in
the $\ct{Path}$ example from \subsecref{Object-Creation}; in that
example $\ct{Path}$ is like a capability object, as it grants the
ability to read files. They are not true capability objects, as other
code can create them `out of thin air' (provided they are allowed
to call the static $\c{\c{\ct{Path}}.\cf{parse}}$ method). This however
does require reasoning over aliasing: we will need to reason what
pre-existing $\ct{Path}$ instances a method might have access to,
such as those accessible in global variables, or in the reachable
object graph of a methods parameter.

A key disadvantage of object-capability systems is that they require
complicated reasoning on object aliasing. For example if a method
has access to an instance of the $\ct{Hashable}$ interface, how can
we be sure that such instance is not also a $\ct{File}$? In contrast,
in $\name$, a method that is not declared with a sufficiently powerful
effect would not be able to call I/O methods on such an object, but
it might be able to call a $\cf{hash}$ function on it. This works
because our effect system restricts what methods can be \emph{called},
not what objects can be accessed or passed around.

The atypical object-capability language of `pop' \cite{Static_use-based_object_confinement}
follows an approach similar to conventional accessibility: the $\ct{File}$
class would declare what code can call its methods. Thus their approach
inherits the problems of accessibility control, the $\ct{File}$ class
would need to be modified if new code wishes to directly call its
methods, even worse, it allows `rights amplification' by code that
has permission to call a certain method on an object. For example,
such code can create a trivial wrapper around a method that can then
be passed to and used by arbitrary code. In contrast, $\name$ is
designed to not have such problems by being the \emph{inverse} of
accessibility control: code declares (in its $\ck{effect}$ annotation)
what methods it has access to call.

Finally, object capabilities usually require explicitly passing objects
to code that needs their capabilities. In particular, to create restricted
effects (such as a $\ct{ConsolePrint}$ object over a $\ct{Console}$
one), a wrapper object needs to be explicitly created, whereas $\name$'s
type system does not require any such input from the programmer. However,
such problems with object-capabilities can be alleviated by using
implicit parameters and/or variables which are kept in scope for entire
modules \cite{A_Capability-Based_Module_System_for_Authority_Control},
this is a simpler alternative to effect systems \cite{Capabilities:_Effects_for_Free},
but is weaker and less flexible.

\pagebreak{}

To summarise: when reasoning over \emph{statically} known behaviour,
$\name$ is less verbose and easier to reason with than object-capabilities;
however $\name$ allows code similar to the object-capability systems
(as in the $\ct{Path}$ example) enabling \emph{dynamic }behaviour.

\section{Conclusion}\label{sec:Conclusion}

In conclusion, $\name$ presents a new and interesting way of statically
reasoning over effectful operation in object-oriented programs. We
have illustrated how it can provide useful guarantees without being
either too strict, or requiring heavy annotation. We have also shown
how our concept of method call effects combine nicely with object-oriented
polymorphism. Though we have presented a formalised version of $\name$,
we have only informally stated and proved its soundness.

We believe that the concept of method call effects is a promising
addition to an OO and imperative language, offering an alternative
to the more common concepts of object-capabilities and monads. In
particular our $\rname$ rule aims to allow the system to be both
flexible and easy to use.

Though we have shown three potential extension to $\name$, there
are more that could be added to $\name$, in particular, allowing
constructors, field operations, and other language primitives to be
used as an effect could be quite useful; however the resulting language
formalism would not be more interesting than the one presented here.
Adding some kind of effect alias or other similar mechanism to enable
reusing effect annotations would also be useful, as would inference
of effect annotations. However with generics and mutually recursive
methods, it is not clear what effects an inference system should choose,
or even whether this is decidable.
\begin{acks}
The authors would also like to thank the anonymous reviewers for their
valuable comments and helpful suggestions. This work is supported
in part by the \grantsponsor{Marsden Fund}{Royal Society of New Zealand (Te Ap\={a}rangi) Marsden Fund (Te P\={u}tea
Rangahau a Marsden)}{https://royalsociety.org.nz/what-we-do/funds-and-opportunities/marsden/}{} under grant \grantnum{Marsden Fund}{VUW1815}{}.
\end{acks}

\appendix
\clearpage{}

\section{Additional Typing Judgements}\label{sec:Auxiliary}

\Secref{Formalism} describes $\name$'s abstract syntax, our notational
conventions, and the typing rules for the judgements of form $\DL\vdash\ensuremath{\ob{\ep}\preceq\ob{\eps '}}$,
$\CC\vdash\SIG\eq\ee$, and $\vdash\ob{\SIG\eq\ee}\vartriangleleft\SIGs '$.
\Figref{Additional-Typing-Judgements} presents the complete set of
typing rules, including the remaining judgements, of forms $\vdash\D$,
$\vdash\SIG$, $\DL\vdash\T$, $\DL\vdash\ep$, that check that declarations,
method signatures, type names, and effects are well typed; as well
as judgements of form $\DL|\GG\vdash\ee:\T|\ob{\ep}$ and $\DL\vdash\T\leq\Ts '$,
which check that an expression has type $\T$ and effects $\ob{\ep}$,
and that $\T$ is a subtype of $\Ts '$. Note that the occurrence
of $\ob{\ob{\vdash\ob{\SIG\eq\ee}\vartriangleleft\SIGs '}}$ in the
$\rn{\vdash\textrm{\ensuremath{\k{class}}}}$ rule is equivalent to
$\vdash\ob{\SIG\eq\ee}\vartriangleleft\SIGs{'_{1,1}},\ldots,\vdash\ob{\SIG\eq\ee}\vartriangleleft\SIGs{'_{n,m}}$,
this is because $\SIG$ and $\ee$ also occur under one overbar (in
the conclusion of the rule), so our notation rule described in \subsecref{Expression-Typing}
applies.\null\vspace{1\baselineskip}

\noindent 

\noindent {\scriptsize{}}%
\noindent\begin{minipage}[t]{1\textwidth}%
{\scriptsize{}$\rc{\vdash\D}{\nir{\vdash\textrm{\ensuremath{\k{class}}}}{\begin{array}{c}
\ob{\SI(\CCs{''})=\ck{interface}\,\CCs{''}\,\br{\ob{\SIGs '\sc}}}\vsep\\
\{\ob{\CCs '}\}\subseteq\op{dom}(\SI)\hsep\ob{\CC\vdash\SIG\eq\ee}\hsep\ob{\ob{\vdash\ob{\SIG\eq\ee}\vartriangleleft\SIGs '}}
\end{array}}{\vdash\k{class}\,\CCp{\ob{\CCs '\,\ff}}\co\ob{\CCs{''}}\,\br{\ob{\SIG\eq\ee\sc}}}}$\hfill{}$\nir{\vdash\textrm{\ensuremath{\k{class}}}}{\begin{array}{c}
\ob{\SI(\CCs{''})=\ck{interface}\,\CCs{''}\,\br{\ob{\SIGs '\sc}}}\vsep\\
\{\ob{\CCs '}\}\subseteq\op{dom}(\SI)\hsep\ob{\CC\vdash\SIG\eq\ee}\hsep\ob{\ob{\vdash\ob{\SIG\eq\ee}\vartriangleleft\SIGs '}}
\end{array}}{\vdash\k{class}\,\CCp{\ob{\CCs '\,\ff}}\co\ob{\CCs{''}}\,\br{\ob{\SIG\eq\ee\sc}}}$\hfill{}\hfill{}$\nir{\vdash\textrm{\ensuremath{\k{interface}}}}{\ob{\vdash\SIG}}{\vdash\k{interface}\,\CC\,\br{\ob{\SIG\sc}}}$\hfill{}\vrule$\rc{\vdash\SIG}{\nir{\vdash\textrm{\ensuremath{\k{class}}}}{\begin{array}{c}
\ob{\SI(\CCs{''})=\ck{interface}\,\CCs{''}\,\br{\ob{\SIGs '\sc}}}\vsep\\
\{\ob{\CCs '}\}\subseteq\op{dom}(\SI)\hsep\ob{\CC\vdash\SIG\eq\ee}\hsep\ob{\ob{\vdash\ob{\SIG\eq\ee}\vartriangleleft\SIGs '}}
\end{array}}{\vdash\k{class}\,\CCp{\ob{\CCs '\,\ff}}\co\ob{\CCs{''}}\,\br{\ob{\SIG\eq\ee\sc}}}}$\hfill{}$\nir{\vdash\SIG}{\{\ob{\CC}\}\subseteq\op{dom}(\SI)\hsep\DL\vdash\T\hsep\ob{\DL\vdash\Ts '}\hsep\ob{\DL\vdash\ep}}{\vdash\T\,\mmap{\ob{\X\co\CC}}{\ob{\Ts '\,\x}}\,\e{\ob{\ep}}}\DL=\ob{\X\mapsto\CC}$\hfill{}\hbox{}}{\scriptsize\par}

\hrule

{\scriptsize{}$\rc{\CC\vdash\SIG\eq\ee}{\nir{\ob{\SIG\eq\ee}\vartriangleleft\SIGs '}{\begin{array}{c}
(\Ts{_{1}}\,\mmap{\ob{\Xs{_{2}}\co\CCs{_{1}}}}{\ob{\Ts{'_{1}\,\xs{_{1}}}}}\,\e{\ob{\eps{_{1}}}}\_)[\ob{\Xs{_{2}}\coloneqq\Xs{_{1}}}]\in\ob{\SIG\eq\ee}\vsep\\
\DL=\ob{\Xs{_{2}}\mapsto\CCs{_{2}}}\hsep\DL\vdash\ob{\eps{_{1}}}\preceq\ob{\eps{_{2}}}\vsep\\
\DL\vdash\Ts{_{1}}\leq\Ts{_{2}}\hsep\ob{\DL\vdash\Ts{'_{2}}\leq\Ts{'_{1}}}\hsep\ob{\DL\vdash\CCs{_{2}}\leq\CCs{_{1}}}\hsep
\end{array}}{\vdash\ob{\SIG\eq\ee}\vartriangleleft\Ts{_{2}}\,\mmap{\ob{\Xs{_{2}}\co\CCs{_{2}}}}{\ob{\Ts{'_{2}\,\xs{_{2}}}}}\,\e{\ob{\eps{_{2}}}}}}$\hfill{}$\nir{\vdash\SIG\eq\ee}{\begin{array}{c}
\vdash\T\,\mmap{\ob{\X\co\CCs '}}{\ob{\Ts '\,\x}}\,\e{\ob{\ep}}\vsep\\
\ob{\X\mapsto\CCs '}|\ck{this}\mapsto\CC,\ob{\x\mapsto\Ts '}\vdash\ee:\T|\ob{\eps '}
\end{array}}{\CC\vdash\T\,\mmap{\ob{\X\co\CCs '}}{\ob{\Ts '\,\x}}\,\e{\ob{\ep}}\eq\ee}\ob{\eps '}\preceq\ob{\ep}$\hfill{}\vrule$\rc{\vdash\ob{\SIG\eq\ee}\vartriangleleft\SIGs '}{\nir{\ob{\SIG\eq\ee}\vartriangleleft\SIGs '}{\begin{array}{c}
(\Ts{_{1}}\,\mmap{\ob{\Xs{_{2}}\co\CCs{_{1}}}}{\ob{\Ts{'_{1}\,\xs{_{1}}}}}\,\e{\ob{\eps{_{1}}}}\_)[\ob{\Xs{_{2}}\coloneqq\Xs{_{1}}}]\in\ob{\SIG\eq\ee}\vsep\\
\DL=\ob{\Xs{_{2}}\mapsto\CCs{_{2}}}\hsep\DL\vdash\ob{\eps{_{1}}}\preceq\ob{\eps{_{2}}}\vsep\\
\DL\vdash\Ts{_{1}}\leq\Ts{_{2}}\hsep\ob{\DL\vdash\Ts{'_{2}}\leq\Ts{'_{1}}}\hsep\ob{\DL\vdash\CCs{_{2}}\leq\CCs{_{1}}}\hsep
\end{array}}{\vdash\ob{\SIG\eq\ee}\vartriangleleft\Ts{_{2}}\,\mmap{\ob{\Xs{_{2}}\co\CCs{_{2}}}}{\ob{\Ts{'_{2}\,\xs{_{2}}}}}\,\e{\ob{\eps{_{2}}}}}}$\hfill{}$\nir{\ob{\SIG\eq\ee}\vartriangleleft\SIGs '}{\begin{array}{c}
(\Ts{_{1}}\,\mmap{\ob{\Xs{_{2}}\co\CCs{_{1}}}}{\ob{\Ts{'_{1}\,\xs{_{1}}}}}\,\e{\ob{\eps{_{1}}}}\_)[\ob{\Xs{_{2}}\coloneqq\Xs{_{1}}}]\in\ob{\SIG\eq\ee}\vsep\\
\DL=\ob{\Xs{_{2}}\mapsto\CCs{_{2}}}\hsep\DL\vdash\ob{\eps{_{1}}}\preceq\ob{\eps{_{2}}}\vsep\\
\DL\vdash\Ts{_{1}}\leq\Ts{_{2}}\hsep\ob{\DL\vdash\Ts{'_{2}}\leq\Ts{'_{1}}}\hsep\ob{\DL\vdash\CCs{_{2}}\leq\CCs{_{1}}}\hsep
\end{array}}{\vdash\ob{\SIG\eq\ee}\vartriangleleft\Ts{_{2}}\,\mmap{\ob{\Xs{_{2}}\co\CCs{_{2}}}}{\ob{\Ts{'_{2}\,\xs{_{2}}}}}\,\e{\ob{\eps{_{2}}}}}$\hfill{}\hbox{}}{\scriptsize\par}

\hrule

{\scriptsize{}$\rc{\DL\vdash\T}{\nir{\vdash\ensuremath{\T.\m}}{\ob{\DL\vdash\Ts '\leq\CCs '}}{\DL\vdash\T.\mma{\ob{\Ts '}}}\ensuremath{\SI\bigl(\DL(\T)\,.\mm\bigr)=\_\,\mmap{\ob{\X\co\CCs '}}{\_}\,\_\sc}}$\hfill{}$\nir{\vdash\T}{\DL(\T)\in\op{dom}(\SI)}{\DL\vdash\T}$\hfill{}\vrule$\rc{\DL\vdash\ep}{\nir{\vdash\ensuremath{\T.\m}}{\ob{\DL\vdash\Ts '\leq\CCs '}}{\DL\vdash\T.\mma{\ob{\Ts '}}}\ensuremath{\SI\bigl(\DL(\T)\,.\mm\bigr)=\_\,\mmap{\ob{\X\co\CCs '}}{\_}\,\_\sc}}$\hfill{}$\nir{\vdash\ensuremath{\T.\m}}{\ob{\DL\vdash\Ts '\leq\CCs '}}{\DL\vdash\T.\mma{\ob{\Ts '}}}\SI\bigl(\DL(\T)\,.\mm\bigr)=\_\,\mmap{\ob{\X\co\CCs '}}{\_}\,\_\sc$\hfill{}\hfill{}$\nir{\vdash\*}{\emptytex}{\DL\vdash\*}$\hfill{}\vrule$\rc{\DL|\GG\vdash\ee:\T|\ob{\ep}}{\nir{\vdash\ensuremath{\T.\m}}{\ob{\DL\vdash\Ts '\leq\CCs '}}{\DL\vdash\T.\mma{\ob{\Ts '}}}\ensuremath{\SI\bigl(\DL(\T)\,.\mm\bigr)=\_\,\mmap{\ob{\X\co\CCs '}}{\_}\,\_\sc}}$\hfill{}$\nir{\vdash\x\cor\T}{\GG(\x)=\T}{\DL|\GG\vdash\x:\T|\emp}$\hfill{}\hbox{}}{\scriptsize\par}

{\scriptsize{}\lrule{3}{1}{$\rc{\DL\vdash\T}{\nir{\vdash\ensuremath{\T.\m}}{\ob{\DL\vdash\Ts '\leq\CCs '}}{\DL\vdash\T.\mma{\ob{\Ts '}}}\ensuremath{\SI\bigl(\DL(\T)\,.\mm\bigr)=\_\,\mmap{\ob{\X\co\CCs '}}{\_}\,\_\sc}}$$\nir{\vdash\T}{\DL(\T)\in\op{dom}(\SI)}{\DL\vdash\T}$$\nir{\vdash\ensuremath{\T.\m}}{\ob{\DL\vdash\Ts '\leq\CCs '}}{\DL\vdash\T.\mma{\ob{\Ts '}}}\SI\bigl(\DL(\T)\,.\mm\bigr)=\_\,\mmap{\ob{\X\co\CCs '}}{\_}\,\_\sc$$\nir{\vdash\*}{\emptytex}{\DL\vdash\*}$}{$\nir{\vdash\x\cor\T}{\GG(\x)=\T}{\DL|\GG\vdash\x:\T|\emp}$}}{\scriptsize\par}

{\scriptsize{}\hfill{}$\nir{\vdash\textrm{\ensuremath{\ck{new}}}}{\begin{array}{c}
\SI(\CC)=\ck{class}\,\CCp{\ob{\T\,\ff}}\,\_\,\br{\_}\vsep\\
\ob{\DL|\GG\vdash\ee:\T|\ob{\ep}}
\end{array}}{\DL|\GG\vdash\ck{new}\,\CCp{\ob{\ee}}:\CC|\ob{\ob{\ep}}}$\hfill{}\hfill{}$\frc{\nir{\vdash\ensuremath{\ensuremath{\ee.\m}}}{\begin{array}{c}
\SI(\DL(\T).\m)=\Ts{''}\,\mmap{\_}{\ob{\Ts '\,\x}}\,\_\sc\vsep\\
\DL|\GG\vdash\ee:\T|\ob{\ep}\hsep\ob{\DL|\GG\vdash\ees ':\Ts '|\ob{\eps '}}
\end{array}}{\DL|\GG\vdash\ee.\mp{\ob{\ees '}}:\Ts{''}|\ob{\ep},\ob{\ob{\eps '}},\T.\m}}$\hfill{}\hfill{}$\nir{\vdash\ensuremath{\ee.\ff}}{\begin{array}{c}
\SI(\CC)=\ck{class}\,\CCp{\ob{\T\,\ff}}\,\_\,\br{\_}\vsep\\
\DL|\GG\vdash\ee:\CC|\ob{\ep}
\end{array}}{\DL|\GG\vdash\ee.\ff:\T|\ob{\ep}}$\hfill{}\hfill{}$\nir{\vdash\ensuremath{\ensuremath{\ee.\m}}}{\begin{array}{c}
\SI(\DL(\T)\,.\m)=\Ts{''}\,\mmap{\_}{\ob{\Ts '\,\x}}\,\_\vsep\\
\DL|\GG\vdash\ee:\T|\ob{\ep}\hsep\ob{\DL|\GG\vdash\ees ':\Ts '|\ob{\eps '}}
\end{array}}{\DL|\GG\vdash\ee.\mp{\ob{\ees '}}:\Ts{''}|\ob{\ep},\ob{\ob{\eps '}},\T.\m}$\hfill{}\hbox{}}{\scriptsize\par}

\vspace{-0.5\baselineskip}

{\scriptsize{}\rrule{2}{2}{$\nir{\vdash\textrm{\ensuremath{\ck{restrict}}}}{\DL|\GG\vdash\ee:\T|\ob{\eps '}\hsep\DL\vdash\ob{\eps '}\preceq\ob{\ep}}{\DL|\GG\vdash\re{\ob{\ep}}\,\ee:\T|\ob{\eps '}}$$\nir{\vdash Sub}{\DL|\GG\vdash\ee:\T|\ob{\ep}\hsep\DL\vdash\T\leq\Ts '}{\DL|\GG\vdash\ee:\Ts '|\ob{\ep}}$}{$\rc{\DL\vdash\T\leq\Ts '}{\nir{\leq\T}{\begin{array}{c}
\SI(\DL(\T))=\ck{class}\,\CCp{\_}\co\ob{\CCs{''}}\,\_\,\br{\_}\vsep\\
\CCs '\in\ob{\CCs{''}},\DL(\T)
\end{array}}{\DL\vdash\T\leq\CCs '}}$$\nir{\leq Refl}{\emptytex}{\DL\vdash\T\leq\T}$$\nir{\leq\T}{\begin{array}{c}
\SI(\DL(\T))=\ck{class}\,\CCp{\_}\co\ob{\CCs{''}}\,\_\,\br{\_}\vsep\\
\CCs '\in\ob{\CCs{''}},\DL(\T)
\end{array}}{\DL\vdash\T\leq\CCs '}$\hfill{}}}{\scriptsize\par}

{\scriptsize{}\hbox{}\hfill{}$\nir{\vdash\textrm{\ensuremath{\ck{restrict}}}}{\DL|\GG\vdash\ee:\T|\ob{\eps '}\hsep\DL\vdash\ob{\eps '}\preceq\ob{\ep}}{\DL|\GG\vdash\re{\ob{\ep}}\,\ee:\T|\ob{\eps '}}$\hfill{}\hfill{}$\nir{\vdash Sub}{\DL|\GG\vdash\ee:\T|\ob{\ep}\hsep\DL\vdash\T\leq\Ts '}{\DL|\GG\vdash\ee:\Ts '|\ob{\ep}}$\hfill{}\vrule$\rc{\DL\vdash\T\leq\Ts '}{\nir{\leq\T}{\begin{array}{c}
\SI(\DL(\T))=\ck{class}\,\CCp{\_}\co\ob{\CCs{''}}\,\_\,\br{\_}\vsep\\
\CCs '\in\ob{\CCs{''}},\DL(\T)
\end{array}}{\DL\vdash\T\leq\CCs '}}$\hfill{}$\nir{\leq Refl}{\emptytex}{\DL\vdash\T\leq\T}$\hfill{}\hfill{}$\nir{\leq\T}{\begin{array}{c}
\SI(\DL(\T))=\ck{class}\,\CCp{\_}\co\ob{\CCs{''}}\,\_\,\br{\_}\vsep\\
\CCs '\in\ob{\CCs{''}},\DL(\T)
\end{array}}{\DL\vdash\T\leq\CCs '}$\hfill{}\hbox{}}{\scriptsize\par}

\hrule

{\footnotesize{}$\rc{\DL\vdash\ensuremath{\ob{\ep}\preceq\ob{\eps '}}}{\nir{\preceq Com}{\DL\vdash\ob{\eps{_{1}}}\preceq\ob{\eps{'_{1}}}\hsep\DL\vdash\ob{\eps{_{2}}}\preceq\ob{\eps{'_{2}}}}{\DL\vdash\ob{\eps{_{1}}},\ob{\eps{_{2}}}\preceq\DL\vdash\ob{\eps{'_{1}}},\ob{\eps{'_{2}}}}}$\hfill{}$\nir{\preceq Ref}{\bigl\{\ob{\ep}\bigr\}=\bigl\{\ob{\eps '}\bigr\}}{\DL\vdash\ob{\ep}\preceq\ob{\eps '}}$\hfill{}\hfill{}$\nir{\preceq Trans}{\DL\vdash\overline{\ep}\preceq\overline{\eps '}\hsep\DL\vdash\ob{\eps '}\preceq\overline{\eps{''}}}{\DL\vdash\ob{\ep}\preceq\ob{\eps{''}}}$\hfill{}\hfill{}$\nir{\preceq Comp}{\DL\vdash\ob{\eps{_{1}}}\preceq\ob{\eps{'_{1}}}\hsep\DL\vdash\ob{\eps{_{2}}}\preceq\ob{\eps{'_{2}}}}{\DL\vdash\ob{\eps{_{1}}},\ob{\eps{_{2}}}\preceq\DL\vdash\ob{\eps{'_{1}}},\ob{\eps{'_{2}}}}$\hfill{}\hbox{}}{\footnotesize\par}

{\footnotesize{}\hbox{}\hfill{}$\nir{\preceq\T}{\DL\vdash\T\leq\Ts '}{\DL\vdash\T.\m\preceq\Ts '.\m}$\hfill{}\hfill{}$\nir{\preceq\ensuremath{\CC.\m}}{\SI(\CC.\m)=\_\,\mmap{\_}{\_}\,\e{\ob{\ep}}\,\_\sc}{\DL\vdash\CC.\m\preceq\ob{\ep}}$\hfill{}\hfill{}$\nir{\preceq\*}{\emptytex}{\DL\vdash\overline{\ep}\preceq\*}$\hfill{}\hfill{}$\nir{\preceq\ensuremath{\emp}}{\emptytex}{\DL\vdash\emp\preceq\overline{\ep}}$\hfill{}\hbox{}}{\footnotesize\par}

\captionof{figure}{\label{fig:Additional-Typing-Judgements}Complete
Set of Typing Judgements}%
\end{minipage}\clearpage{}

\bibliographystyle{ACM-Reference-Format}
\bibliography{call-effects}

\end{document}